\documentclass[twoside,11pt]{article}

\usepackage[preprint]{jmlr2e}
\hypersetup{colorlinks=true,urlcolor=blue, citecolor=blue}
\usepackage{dirtytalk}

\usepackage{multirow}

\include{macro_jmlr}

\jmlrheading{}{}{}{April 21, 2021}{}{}{Ali Mahzarnia and Jun Song}

\ShortHeadings{functional predictor selection}{Mahzarnia and Song}
\firstpageno{1}

\begin{document}

\title{Multivariate functional group sparse regression: functional predictor selection}

\author{\name Ali Mahzarnia \email amahzarn@uncc.edu \\
       \addr Department of Mathematics and Statistics\\
       University of North Carolina at Charlotte\\
       Charlotte, NC 28223, USA
       \AND
       \name Jun Song\thanks{
    Corresponding author.}\email Jun.Song@uncc.edu \\
       \addr Department of Mathematics and Statistics\\
       University of North Carolina at Charlotte\\
       Charlotte, NC 28223, USA}

\editor{}

\maketitle

\begin{abstract}
In this paper, we propose methods for functional predictor selection and the estimation of smooth functional coefficients simultaneously in a scalar-on-function regression problem under high-dimensional multivariate functional data setting. In particular, we develop two methods for functional group-sparse regression under a generic Hilbert space of infinite dimension. We show the convergence of algorithms and the consistency of the estimation and the  selection (oracle property) under infinite-dimensional Hilbert spaces. Simulation studies show the effectiveness of the methods in both the selection and the estimation of functional coefficients. The applications to the functional magnetic resonance imaging (fMRI) reveal the regions of the human brain related to ADHD and IQ. 
\end{abstract}

\begin{keywords}
  functional predictor selection, multivariate functional group lasso, oracle property, fMRI, ADHD, IQ
\end{keywords}

\section{Introduction}
In the past decades, functional data analysis (FDA) has received a great attention in which an entire function is an observation. \cite{Ramsay2005} introduced a general framework of FDA and many other researchers investigated the estimation and inference methods of functional data. See \cite{Yao2005}, \cite{Yao2005a}, \cite{Horvath2012}, and \cite{wang2016functional}. More recently, FDA has been extended to multivariate functional data that can deal with multiple functions as a single observation. See \citep{CHIOU2016301, happ2018multivariate}. However, the sparseness of functional predictors in the multivariate model has not been studied well compared to the univariate case. Hence, we aim to develop theories and algorithms for the sparse functional regression methods with functional predictor selection when we have scalar data as response values and high-dimensional multivariate functional data as predictors. 

Under the multivariate setting, numerous sparse models have been studied with the introduction of $L_1$-penalty. Least absolute shrinkage and selection operator (LASSO) introduces a penalty term to the least square cost function which performs both variable selection and shrinkage \citet{LASSO}. The LASSO-type penalty, such as the Elastic Net \citet{elast}, the smoothly clipped absolute deviation (SCAD) \citet{SCAD}, their modifications (the adaptive LASSO \citet{zou2006adaptive} and the adaptive Elastic Net \cite{adelas}) are developed to overcome the lack of theoretical support and the practical limitations of the LASSO such as the saturation. These methods were developed to overcome the challenges and enjoy asymptotic properties when the sample size increases, such as the estimation consistency and the selection consistency, also known as the oracle property.  

Recently, the sparse models have been extended to the functional data. Initially, a majority of the literature seeks the sparseness of the time domain. Examples include \cite{james2009functional} and related articles for univariate functional data and \cite{blanquero2019variable} for multivariate functional data. On the other hand, \cite{Robust} proposed a model considering the sparseness in the functional predictors under the multivariate functional data setting. In particular, they introduced a model based on the least absolute deviation (LAD) and the group LASSO in the presence of outliers in functional predictors and responses. Its numerical examples and data application show the effectiveness in practice, but theoretical properties and detailed algorithm have not been explored. To this end, we develop methods for scalar-on-function regression model which allows sparseness of the functional predictors and the simultaneous estimation of the smooth functional coefficients. To implement it with the actual data, we derive two algorithms for each of the optimization problems. Finally, we show both the functional predictor selection consistency and the estimation consistency. 

\begin{figure}[!ht]
    \centering
    \includegraphics[scale=1]{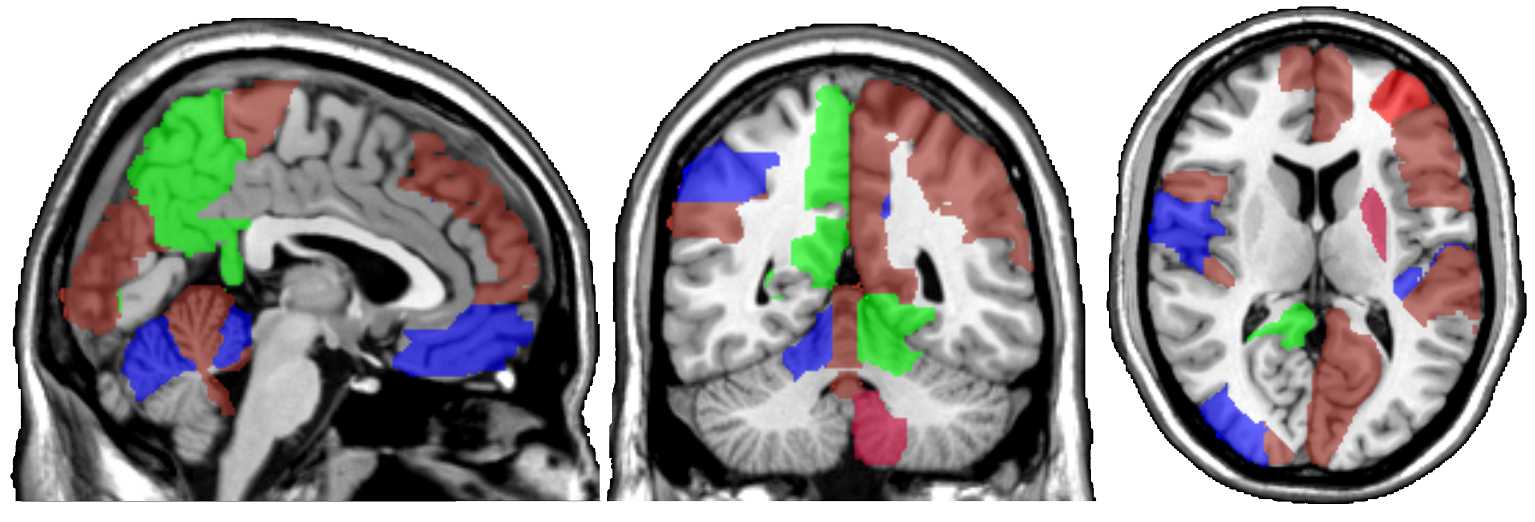}
    \caption{The regions of interests, the BOLD activities of which correlate the most with the ADHD score variability in a sample of subjects and achieve the lowest error in estimation. The regions associated with ADHD are colored red, those associated with ADHD Hyper/Impulsive are blue, and the ones  associated with ADHD Inattentive are colored green.}
    \label{figint}
    \end{figure}

One motivating example for our methods is the application to the functional magnetic resonance imaging (fMRI). The dataset consists of the functional signals of the brain activities measured by blood-oxygen-level-dependent (BOLD), which detects hemodynamic changes based on the metabolic demands followed by neural activities. There are pre-specified regions of the brain, and the BOLD signals associated with multiple voxels in each region are integrated into one signal for that region. Thus, the fMRI data are considered to be multivariate functional data in which each functional predictor represents the signals from a region of the brain. In Section \ref{sec: real data application}, we regress the ADHD index to the regional BOLD activities of the fMRI of the human subjects. There are $116$ regions of the brain in the data, and our methods reduce the regions to 41 regions with significantly lower errors than the linear functional regression. Figure \ref{figint} displays the regions of the brain’s atlas that are identified by our method. It shows that the methods simplify the data analysis and provide clear representation while keeping the crucial information. The analysis shows that there is an urgent need for new methods in the fields of medical and life sciences as well as other related areas.
The following quote from \cite{bandettini2020fmri} further motivates to study the applications of the sparse multivariate functional regression in the field of fMRI.\\

{\small \say{Think of the  challenge of the fMRI with the analogous situation one would have if, when flying over a city at night, an attempt is made to determine the city activities in detail by simply observing where the lights are on. The information is extremely sparse, but with time, specific inferences can be drawn.}}
\rightline{{\rm ---  Peter A. Bandettini , \emph{fMRI}, 2020 \hspace{1.1cm} }}\\ 

The rest of the paper is organized as follows. In Section \ref{sec:preliminary and notation}, we illustrate the general framework of our methods along with the notations used in this paper. In Section \ref{sec: model description}, we describe the model and the optimization problem that we consider. Then, we develop an explicit solution to the optimization problem, and illustrate a detailed procedure using alternating  direction  method  of  multipliers (ADMM) in Section \ref{sec: estimation}. We also derive another algorithm, called groupwise-majorization-descent (GMD), along with the strong rule for faster computation in Section \ref{sec: GMD}.
In Section \ref{sec: asymptotic}, we develop asymptotic results, including the consistency of our methods and the oracle property. In Section \ref{sec: simulation}, we show the effectiveness of the methods by conducting simulation studies. In Section \ref{sec: real data application}, we apply the methods to a resting state fMRI dataset. Concluding discussions are made in section \ref{sec: discussion}. Finally, the appendix includes all of the proofs and the list of the regions of the brain associated with the ADHD and the IQ scores. We created an \texttt{R} package \texttt{MFSGrp} for the computation, and it is available at {\url{https://github.com/Ali-Mahzarnia/MFSGrp}}.

\section{Preliminary and notation}\label{sec:preliminary and notation}
Let $(\Omega, \ca F, P)$ be a probability space. Let $T \lo j$ be a compact set in $\real \hi {d\lo j}$ for $j=1, \ldots, p$. Let $\H \hi 1, \ldots, \H \hi p$ be separable Hilbert spaces of functions from $T \lo j$ to $\real$ with an inner product $\langle \cdot, \cdot \rangle \lo {\H \hi j}$.
Let $\H = \H \hi 1  \times \cdots \times \H \hi p$ be endowed with the inner product
\begin{align*}
\langle f, g \rangle \lo {\H} = \langle f\hi 1 , g\hi 1  \rangle \lo {\H\hi 1} + \ldots + \langle f\hi p, g\hi p \rangle \lo {\H\hi p}, 
\end{align*}
for any $f=(f\hi 1 , \ldots, f\hi p)\trans \in \H$, and $g=(g\hi 1 , \ldots, g \hi p)\trans \in \H$. Then, $\H$ is also a separable Hilbert space. Let $X:\Omega \rightarrow \H$ be a measurable function with respect to $\ca F\lo X / \ca B\lo X$ where $\ca B \lo X$ is the Borel $\sigma$-field generated by open sets in $\H$.

Let $X$ be a random element in $\H$. If $E\|X\|\lo {\H}<\infty$; then, the linear functional $f\mapsto E\inner{f,X}{\H}$ is bounded. By the Riesz's representation theorem, there is a unique element in $\H$, say $\mu \lo X$, such that $\inner{\mu \lo X,f}{\H}= E\inner{f,X}{\H}$ for any $f \in \H$. \cite{Conway1990}. We call $\mu \lo X$ the mean element of $X$ or expectation of $X$. If we can further assume $E\|X\|\hi 2 \lo {\H}<\infty$, the operator $\H \rightarrow \H$,
\begin{align}
    \Gamma\lo {XX} = E[\{X-E(X)\}\otimes \{X-E(X)\}],
\end{align}
exists and is a Hilbert-Schmidt operator, where $\otimes$ indicates a tensor product computed in a way that for $x,y,z\in \H$, $(x\otimes y)(z)=\inner{y,z}{\H}x$. \cite{hsing2015theoretical}. 

Let $Y$ be a random element in $\H \lo Y$. Subsequently, we can define the covariance operator between $X$ and $Y$ by
\begin{align*}
    \Gamma \lo {YX} = E[\{Y-E(Y)\} \otimes \{ X-E(X)\}],
\end{align*}
which maps from $\H$ to $\H \lo Y$. $\Gamma\lo{XY}$ can be similarly defined. For convenience, throughout this paper, we assume that $E(X)=0$ and $E(Y)=0$  without loss of generality. Hence, the regression model is 
\begin{align*}
    Y = \inner{X,\beta}{\H} + \epsilon,
\end{align*}
where $\beta \in \H$ is the unknown coefficient function, and $\epsilon$ is an error term which is a mean zero random variable and independent of $X$. Consider $Y$ as a scalar random variable.  We can rewrite $\beta(\cdot) = (\beta \hi 1(\cdot), \ldots, \beta \hi p(\cdot))$ and   
\begin{align*}
\inner{X,\beta}{\H}=\tsum\lo {j=1}\hi p \inner{X\hi j, \beta \hi j}{\H\hi i}.
\end{align*}

\section{Model description}\label{sec: model description}
We are interested in the situations where the predictors are multivariate functions but only a few functional predictors affecting the response. i.e., a random variable $Y$ and random functions $X\hi j \in \H\hi j$ have the following relation,
\begin{align}\label{eq: pop model}
    Y = \sum\lo {j \in {\stens A}} \inner{X \hi j, \beta \hi j}{\H\hi j} + \epsilon,
\end{align}
where $\sten A \subseteq \{1,\ldots, p\}$ is an unknown active set of indices involved in this regression model, and $\epsilon$ is a mean zero error term that is independent of $X$. 

Assume that we have a random sample of size $n$ from the model (\ref{eq: pop model}). To estimate $\beta$ and the active set $\sten A$, we propose the following objective function.

\begin{align} \label{eq: sam model objective-1}
L(\beta;\lambda\lo {1n}) = \frac{1}{2}E\lo n(Y -\inner{X,\beta}{\H} )^2 + \lambda\lo {1n} \tsum\lo {j=1}\hi p \| \beta\hi j\|\lo {\H\hi j} , \quad \beta \in \H,
\end{align}
where $E\lo n$ is the expectation with the empirical distribution. We added the group-lasso type penalty so that each group includes one functional component in the infinite dimensional Hilbert space, $\H^j$, $j=1,\ldots, p$. Note that the norm in the penalty term is $L\lo 2$-norm which makes the objective function convex. In addition, we propose an alternative objective function to gain a more stable solution path.

\begin{align} \label{eq: sam model objective-2}
L(\beta;\lambda\lo {1n}, \lambda \lo {2n}) = \frac{1}{2}E\lo n(Y -\inner{X,\beta}{\H} )^2 + \lambda\lo {1n} \tsum\lo {j=1}\hi p \| \beta\hi j\|\lo {\H\hi j} + \lambda\lo {2n} \tsum\lo {j=1}\hi p \| \beta\hi j\|\lo {\H\hi j} \hi 2 , \quad \beta \in \H,
\end{align}
The quadratic term allows us to have a stable solution path and encourages further grouping effects. It is similar to the Elastic Net proposed by \cite{elast}, but it is different in that the norm in the first penalty term uses $L\lo 2$-norm, and both the two penalties are applied group-wisely. The group-wise second penalty also gives us a huge computational advantage. 

Furthermore, we also consider the smoothing penalty of the functional coefficients $\beta \in \H$ by adding the term, $\lambda_{3n} \|\beta''\|^2\lo {\H} $ to the objective functions, (\ref{eq: sam model objective-1}) and (\ref{eq: sam model objective-2}). It allows us to estimate smooth functional coefficients and to select the functional predictors simultaneously. In addition, it provides a better interpretation of the functional coefficients in this linear functional regression model.

\section{Estimation: ADMM}\label{sec: estimation}
In this section, we develop the algorithm for solving the optimization problems introduced in Section \ref{sec: model description} via the alternating direction method of multipliers (ADMM), one that is popularly used in a general convex optimization problem. See \cite{ADMM}. Consider the following optimization problem.
\begin{align}\label{eq: original lagrange objective function }
\arg\min_{\beta,\gamma} \quad & {f} (\beta)+{g}(\gamma)\\
\textrm{s.t.} \quad  & \beta-\gamma=0, \nonumber
\end{align}
where $\gamma$ is duplicate variable in $\H$,  $ f(\beta)=\frac{1}{2}E_n(Y-\langle X, \beta \rangle \lo {\H}  )^2$,  and $g(\gamma)= \lambda  \sum_{j=1}^p \| \gamma^j\|_{\H^j} $.  Blocks  $ \gamma^j $ are associated with their counterparts' blocks $ \beta^j $. The augmented Lagrangian with its parameter $\rho>0$ is 
\begin{align} \label{lagr}
L_{\rho}(\beta,\gamma, \eta)= f(\beta)+g(\gamma)+ \langle \eta , \beta-\gamma \rangle \lo {\H} +\frac{\rho}{2} \| \beta - \gamma \|^2_{H},
    \end{align}
where the Lagrangian multiplier is $\eta \in \H$. The ADMM update rules are
\begin{align} \label{eq: admm update-population-non-scaled}
\begin{split}
\beta\hi {\text{new}}:&=\arg\min_{\beta} L_{\rho}(\beta,\gamma, \eta)\\
\gamma\hi {\text{new}}:&=\arg\min_{\gamma} L_{\rho}(\beta\hi {\text{new}},\gamma,\eta)\\
\eta\hi {\text{new}}:&=\eta+ \rho (\beta\hi {\text{new}}-\gamma\hi {\text{new}}).
\end{split}
\end{align}
For computational convenience, it is a usual practice to consider the scaled dual parameter of the ADMM. Let $u=\frac{1}{\rho} \eta$. It is straightforward to verify that the update rules (\ref{eq: admm update-population-non-scaled}) with scaled dual parameter are equivalent to
\begin{align} \label{eq: admm update-population-scaled}
\begin{split}
\beta\hi {\text{new}}:&=\arg\min_{\beta} \left(  f(\beta)+\frac{\rho}{2}\|\beta-\gamma+ U\|^2_{\H} \right) \\
\gamma\hi {\text{new}}:&=\arg\min_{\gamma}  \left( g(\gamma)+\frac{\rho}{2}\|\beta\hi {\text{new}}-\gamma+ U\|^2_{\H} \right)\\
U\hi {\text{new}}:&=U+ \beta\hi {\text{new}}-\gamma\hi {\text{new}}.
\end{split}
\end{align}

\subsection{Coordinate representation of functional data}\label{sec: coordinate representation}
Our method is based on the basis-expansion approach to the functional data. Suppose that we have $n$ random copies from the model (\ref{eq: pop model}) denoted by $(X\lo 1, Y \lo 1), \ldots, (X\lo n, Y \lo n)$ and we observe $X \lo i\hi j$ on $\{ t\hi j\lo {i1}, \ldots, t\hi j\lo {i{a\lo i \hi j}}\}$ for each $i=1,\ldots,n$ and $j=1,\ldots, p$. 

At the sample level, we assume that $\H \hi j$ is spanned by a given set of basis functions, ${\cal B} \hi j = \{b\lo 1 \hi j, \ldots, b \lo {m\lo j} \hi j\}$. Thus, for any $f \in \H \hi j$, there exist a unique vector $a \in \real \hi {m\lo j}$ such that $f(\cdot)=\sum\lo {k=1}\hi {m\lo j} a\lo k b\lo k \hi j (\cdot)$. We call the vector $a$, the coordinate of $f$ and denote it $[f]\lo {{\cal B}\hi j}$. We also assume that $\H \hi j$ is constructed with the $L\lo 2$-inner product with respect to the Lebesgue measure,
\begin{align*}
\langle f, g \rangle \lo {\H \hi j} = \int_{T\lo j} f(t)g(t) dt, \quad \text{for any }f,g\in \H \hi j.
\end{align*}
Let $G\hi j$ be $m\lo j \times m\lo j$ matrix whose $(i,k)$-th entry is $\inner{b\lo i\hi j, b\lo k\hi j}{\H\hi j}= \int_{T\lo j} b\hi j\lo i(t)b\hi j\lo k(t) dt$, and let $G$ be $M\times M$ block-diagonal matrix whose $j$-th block is $G\hi j$ where $M =\sum\lo {j=1}\hi p m \lo j$. Consequently, for any $f,g \in \H$, 
\begin{align*}
    \langle f, g \rangle \lo {\H} = \tsum\lo {j=1}\hi {p}\tsum \lo {i=1}\hi {m\lo j} \tsum \lo {k=1}\hi {m \lo j}([f\hi j]\lo {{\cal B}\hi j})\lo i ([g\hi j]\lo {{\cal B}\hi j})\lo k\inner{b\lo i\hi j, b\lo k\hi j}{\H\hi j} = \tsum\lo {j=1} \hi p [f\hi j]\lo {{\cal B}\hi j}\trans G \hi j [g\hi j]\lo {{\cal B}\hi j}  = [f]\lo {{\cal B}}\trans G[g]\lo {{\cal B}},
\end{align*}
where $[f]\lo {{\cal B}},[g]\lo {{\cal B}}$ are the $\real \hi M$-dimensional vectors obtained by stacking $[f\hi j]\lo {{\cal B}\hi j}$ and $[g\hi j]\lo {{\cal B}\hi j}$ respectively. We use the basis-expansion approach for each functional covariate $X\lo i \hi j$ for $i=1,\ldots,n$ and $j=1,\ldots, p$, which is also used in \cite{song2021, li-song-2018}. Without loss of generality, we assume $m=m\lo 1 = \cdots = m\lo p$ and $M=pm$.

Suppose that $A$ is a linear operator from $\H\lo 1$ to $\H \lo 2$ in which the basis for $\H \lo 1$ is ${\cal B} = \{ b\lo 1, \ldots, b\lo m\}$ and the basis for $\H \lo 2$ is ${\cal C}= \{c \lo 1, \ldots, c\lo k\}$. Then, we define the coordinate representation of the operator $A$ to be $k \times m$ matrix, say ${}\lo {\cal C}[A]\lo {\cal B}$, whose $(i,j)$-th entry is $([A b\lo j]\lo {\cal C})_i$. It can be easily shown that ${}\lo {\cal C}[Ax]\lo {\cal B}={}\lo {\cal C}[A]\lo {\cal B}[x]\lo {\cal B}$ for any $x \in \H \lo 1$. For notational convenience, if the basis system is obvious in the context, we remove the subscripts of the coordinate representation throughout this paper. The following lemma provides a further simplification for easy computations.

\begin{lemma}\label{lem: coordinate representation covariance operators}
    Let $Q = I - n\inv 1\lo n 1\lo n \trans$. Let $[X\lo {1:n}]\lo {\cal B}$ be the $pm \times n$ matrix, the $k$-th column of which is $[X\lo k]\lo {\cal B}$. Then
    \begin{align*}
        {}\lo {\cal B}[\hat{\Gamma}\lo {XX}]\lo {\cal B} =n\inv [X\lo {1:n}]\lo {\cal B}  Q [X\lo {1:n}]\lo {\cal B}\trans G
        =n\inv {}[\tilde{X}\lo {1:n}]\lo {\cal B}  [\tilde{X}\lo {1:n}]\lo {\cal B}\trans G,
    \end{align*}
    where $[\Xt\lo {1:n}]{}\lo {\cal B}=  [X\lo {1:n}]\lo {\cal B}  Q$. In addition, let $Y$ be the $n$-dimensional vector, the elements of which are the observations $Y_1, \ldots, Y_n$. Then
    \begin{align*}
        [\hat{\Gamma}\lo {YX}] = n\inv Y\trans[\tilde{X}\lo {1:n}]\lo {\cal B} \trans G.
    \end{align*}
\end{lemma}

\subsection{Orthogonalization}
To achieve computational efficiency, we orthonormalize the basis system via Karhunen-Lo\`{e}ve expansion of the covariance operator of each of the functional predictors. For each $j=1, \ldots, p$, define $\Gamma \lo {jj}$ to be the covariance operator of $X \hi j$. Consequently, we have the following lemma.

\begin{lemma}
    Let $(\lambda \hi j \lo  1, v\hi j \lo  1), \ldots, (\lambda \hi j \lo  m, v\hi j \lo  m)$ be the pairs of eigenvalues and vectors of\\ $(G\hi j)\hi {1/2}[\tilde{X}\hi j \lo {1:n}]\lo {{\cal B}\hi j}  [\tilde{X} \hi j\lo {1:n}]\lo {{\cal B}\hi j}\trans (G\hi j)\hi {1/2}$ with $\lambda \hi j \lo  1\ge \ldots \lambda \hi j \lo  m$, and let $[\phi \hi j\lo k]\lo {{\cal B}\hi j}=(G\hi j)\hi {-1/2}v\hi j \lo  k$ for $k=1, \ldots, m$. Then, the Karhunen-Lo\`{e}ve expansion of $\hat{\Gamma}\lo {jj}$ is
    \begin{align*}
        \hat{\Gamma}\lo {jj} = \sum\lo {k=1}\hi m \lambda \hi j \lo  k \phi \hi j \lo  k \otimes \phi \hi j \lo  k.
    \end{align*}
\end{lemma}
Define a $m \times m$ matrix 
\begin{align*}
    \Phi \hi j=  \begin{pmatrix}
    [\phi \hi j \lo  1]\lo {{\cal B}\hi j} & \cdots & [\phi \hi j \lo  m]\lo {{\cal B}\hi j}
    \end{pmatrix}.
\end{align*}
Since $\phi \hi j \lo  m$'s are the eigenfunctions of a self-adjoint operator, they are orthonormal. Thus, for any $x \in \H\hi j$,
\begin{align*}
    x(\cdot) &= \sum\lo {k=1}\hi m \inner{x,\phi \hi j \lo  k}{\H \hi j}\phi \hi j \lo  k(\cdot)\\
    &= \sum\lo {k=1}\hi m [x]\lo {{\cal B}\hi j}\trans G\hi j [\phi \hi j \lo  k] \lo {{\cal B}\hi j} \phi \hi j \lo  k(\cdot).\\
\end{align*}
Define ${\cal C}\hi j = \{\phi \hi j \lo 1, \ldots, \phi \hi j \lo  m\}$ to be the new basis system for $\H\hi j$. Then, we have
\begin{align*}
    [X\hi j \lo {i}]\lo {{\cal C}^j} = (\Phi^j )\trans G^j [X\hi j\lo i]\lo {{\cal B}\hi j}, \quad i=1,\ldots, n, j=1, \ldots, p.
\end{align*}
We assume that the coordinate of $\H$ is based on the orthonormal basis system throughout this section. Thus,
\begin{align*}
    [\hat{\Gamma}\lo {XX}]= \text{diag}(\lambda \lo 1 \hi 1, \ldots, \lambda \lo m \hi 1, \ldots, \lambda \lo 1 \hi p, \ldots, \lambda \lo m \hi p), \quad [X\lo {1:n}] = \text{diag}(\Phi\lo 1 \trans, \ldots, \Phi \lo p \trans) G [X\lo {1:n}]\lo {{\cal B}}, 
\end{align*}
and $\inner{f,g}{\H}=[f]\trans [g]$ for any $f,g\in \H$. 

\subsection{Estimation}\label{subsec: estimation}
Using the representation, we can express the optimization (\ref{eq: admm update-population-scaled}) as follows.
\begin{align} 
\begin{split}\label{eq: admm update-coordinate-scaled}
[\beta\hi {\text{new}}]:&=\arg\min_{\beta\in\H} \left(  {f}(\beta)+\frac{\rho}{2}([\beta]-[\gamma]+ [U])\trans ([\beta]-[\gamma]+ [U]) \right) \\
[\gamma\hi {\text{new}}]:&=\arg\min_{\gamma\in\H}  \left( {g}(\gamma)+\frac{\rho}{2}([\beta\hi {\text{new}}]-[\gamma]+ [U])\trans([\beta\hi {\text{new}}]-[\gamma]+ [U]) \right)\\
[U\hi {\text{new}}]:&=[U]+ [\beta\hi {\text{new}}]-[\gamma\hi {\text{new}}],
\end{split}
\end{align}
where 
\begin{align*}
    f(\beta) &= \frac{1}{2}E\lo n(Y -\inner{X,\beta}{\H} )^2
    =(2n)\inv \tsum\lo {i=1} \hi n \{Y \lo i \hi 2 - 2 (Y\lo i \otimes X\lo i)\beta + \inner{ \beta, (X\lo i \otimes X \lo i)\beta}{\H}\}\\
    &=\frac{1}{2}\hat{\sigma}\lo {YY} -  \hat{\Gamma}\lo {YX}\beta + \frac{1}{2}\inner{\beta,\hat{\Gamma}\lo {XX} \beta}{\H}
    =\frac{1}{2}\hat{\sigma}\lo {YY} - [\hat{\Gamma}\lo {YX}][\beta] + \frac{1}{2}[\beta]\trans[\hat{\Gamma}\lo {XX}] [\beta],
    \end{align*} 
    and $g(\gamma)=\lambda \tsum\lo {j=1}\hi p \| \gamma\hi j\|\lo {\H\hi j} =\lambda \tsum\lo {j=1}\hi p \sqrt{[\gamma\hi j]\trans [\gamma\hi j]}$.

Under the finite-dimensional representation of functional element in $\H$, one can see that the optimization in (\ref{eq: admm update-coordinate-scaled}) is a convex optimization problem.  

\begin{theorem}\label{thm: update admm group LASSO} 
The solution to the optimization problem (\ref{eq: sam model objective-1}) can be achieved by iterating over the following update rules.
\begin{align} \label{eq: update admm group LASSO}
&[\beta\hi {\text{new}}]=([\tilde{X}\lo {1:n}] [\tilde{X}\lo {1:n}]\trans + n\rho I\lo M)^{-1}([\tilde{X}\lo {1:n}] Y+ n\rho([\gamma]- [U]))& \nonumber\\
&[(\gamma^j)\hi {\text{new}}]= S^{\H^j}_{\frac{\lambda}{\rho}}([(\beta^j)\hi {\text{new}}]+[U^j]) & j=1 \dots p\\
&[U\hi {\text{new}}]=[U]+ [\beta\hi {\text{new}}]-[\gamma\hi {\text{new}}],\nonumber &
\end{align}
where $[\gamma\hi j]$, $[U\hi j]$ are corresponding blocks to $[\beta\hi j]$, and $S^{\H^j}_{\lambda} (h)=1_{ \{\| h\|\lo {\H\hi j} > \lambda \}}\left(1-\frac{\lambda}{\| h\|\lo {\H\hi j}}\right)_{+}h$ for $h\in \H \hi j$.
\end{theorem}
If we do not consider orthogonalization, Theorem \ref{thm: update admm group LASSO} would contain element $G \hi j$ in the updates. In this case, the proof of numerical convergence of the update rules is slightly different from that of \cite{ADMM}. However, due to the orthogonalization, the proof of the numerical convergence of the updates in the Theorem \ref{thm: update admm group LASSO} to the solution of the optimization problem (\ref{eq: sam model objective-1}) is identical to that of the ADMM in \cite{ADMM}. Hence, it is omitted.  

\subsection{Different penalty terms} \label{Different penalty section}
In this section, we investigate the different penalty terms in two directions: one for the functional predictor selection, and the other one for the smooth coefficient functions $\beta$. 

\subsubsection{Multivariate Functional Group Elastic Net}
 LASSO does not provide a unique solution. In order to achieve uniqueness and overcome the saturation property, Elastic Net penalty has been introduced by combining the $\ell\lo 1$-norm and $\ell\lo 2$-norm by \cite{elast} for the multivariate data. Functional data are intrinsically an infinite-dimensional objects. Thus, we propose a multivariate functional-version optimization problem for the Elastic net penalty by grouping each functional predictor as follows.
\begin{align}\label{eq: objective function elastic net}
\frac{1}{2}E_n(Y -\inner{X,\beta}{\H} )^2 + \lambda (1-\alpha) \tsum\lo {j=1}\hi p \| \beta\hi j\|\lo {\H\hi j} + \alpha \lambda\tsum\lo {j=1}\hi p \| \beta\hi j\|^2\lo {\H\hi j},
\end{align}
where $\alpha \in [0,1]$ and $\lambda >0$ are the tuning parameters.

This optimization problem still follows the structure of the ADMM algorithm in (\ref{eq: original lagrange objective function }) with $g(\gamma)=\lambda (1-\alpha) \tsum\lo {j=1}\hi p \| \gamma \hi j\|\lo {\H\hi j} + \alpha\lambda \tsum\lo {j=1}\hi p \| \gamma \hi j\|^2\lo {\H\hi j} $. It can be easily shown that the only difference from the original version is the $\gamma$-update in Theorem \ref{thm: update admm group LASSO}. Hence, we have the following update rules.
\begin{theorem}\label{thm: elastic net}
    The solution to the optimization problem (\ref{eq: objective function elastic net}) can be achieved by iterating over the following update rules.
    \begin{align} \label{eq: gamma update elastic net}
&[\beta\hi {\text{new}}]=([\tilde{X}\lo {1:n}] [\tilde{X}\lo {1:n}]\trans + n\rho I\lo M)^{-1}([\tilde{X}\lo {1:n}] Y+ n\rho([\gamma]- [U]))& \nonumber\\
        &[(\gamma^j)\hi {\text{new}}]= \frac{\rho}{\rho + 2\alpha\lambda}S^{\H^j}_{\frac{\lambda (1-\alpha)}{\rho}}([(\beta^j)\hi {\text{new}}]+[U^j])  & j=1 \dots p\\
        &[U\hi {\text{new}}]=[U]+ [\beta\hi {\text{new}}]-[\gamma\hi {\text{new}}].\nonumber &
    \end{align}
\end{theorem}
Regularization parameters can be adjusted through a net search cross validation.

\subsubsection{Smoothness of functional coefficients $\beta$}
According to the simulation, we found that the previous algorithm provides wiggly estimation of functional coefficients $\beta$ most of the time. It might be fine if we are only interested in the prediction; however, it is not the case, because we consider the linear functional regression.We propose an algorithm which controls the roughness of $\beta$ simultaneously to avoid the over-fitting problems and to obtain smooth functional coefficients. In particular, we impose the penalty on the curvature of the coefficients by adding $\frac{\lambda_\text{der}}{2} \|\beta''\|^2\lo \H$ to the objective function (\ref{eq: admm update-population-scaled}). We include this term in $f(\cdot)$ function in the ADMM structure. Finally, the first update rule (\ref{eq: update admm group LASSO}) in Theorem \ref{thm: update admm group LASSO} becomes
\begin{align} \label{secder}
    [\beta\hi {\text{new}}]:=([\tilde{X}\lo {1:n}] [\tilde{X}\lo {1:n}]\trans + n\rho I_M+ \lambda_{\text{der}}  {G''})^{-1}([\tilde{X}\lo {1:n}] Y+ n\rho([\gamma]- [U])), 
\end{align}
where $G''$ is a block-diagonal matrix whose $j$-th block matrix is $((G \hi j)'')\lo {ik}=\int\lo {T \hi j} (\phi\hi j\lo i)''(t)(\phi\hi j\lo k)''(t)dt = \inner{(\phi\hi j \lo i)'',  (\phi\hi j \lo k)''}{\H\hi j} $ for $i,k=1, \ldots, m$, $j=1\ldots, p$. 

For each $j$, $(G^j)''$ can be derived from the second derivative Gram matrix for the original basis, say $(B^j)''$, where $((B^j)'')_{ik} = \int\lo {T \hi j} (b\hi j\lo i)''(t)(b\hi j\lo k)''(t)dt = \inner{(b\hi j \lo i)'',  (b\hi j \lo k)''}{\H\hi j}$. Note that
\begin{align*}
    [\phi^j_{i}]_{{\cal B}^j} = (G^j)\inv ((\Phi^j)\inv )\trans [\phi^j_{i}]_{{\cal C}^j} = (G^j)\inv ((\Phi^j)\inv )\trans e_i,
\end{align*}
where $e_i$ is $i$-th standard basis in $\real^m$. Then,
\begin{align*}
    \inner{(\phi\hi j \lo i)'',  (\phi\hi j \lo k)''}{\H\hi j} 
    &= \inner{\tsum\lo {\ell=1}\hi m ([\phi \lo i \hi j]\lo {{\cal B}\hi j})\lo {\ell} (b\lo \ell \hi j)'', \tsum\lo {\ell=1}\hi m ([\phi _k \hi j]\lo {{\cal B}\hi j})\lo {\ell} (b\lo \ell \hi j)''}{\H \hi j} \\
    &= e_i \trans (\Phi^j)\inv (G^j)\inv (B^j)'' (G^j)\inv (\Phi^j)\inv )\trans e_j.
\end{align*}
where $((B\hi j)'')_{ik} = \int (b \hi j \lo i)''(t)(b \hi j \lo k)''(t)dt$. 

\subsubsection{Tuning}
The initial values for $\gamma$ and $U$ are zero, and the initial $\beta$ is the ridge regression estimation in the first update rule (\ref{eq: update admm group LASSO}). We set the augmented parameter, or the step size, $\rho$ to be $1$ and stay the same through the algorithm. The different values of $\rho$ only change the values of the optimal $\lambda$ on the grid or optimal $(1-\alpha)\lambda$ on the net. The larger the $\rho$, the smaller the optimized regularization parameter of the soft threshold operator. In some practices of augmented Lagrangian, it is possible to choose a small step size and increase it to $1$ gradually in each iteration. It is also stated in \citet{ADMM} why $\rho=1$ is a suitable choice in the ADMM algorithm.   

We use the k-fold cross validation for choosing the mixing parameter $\alpha$, regularization parameter of the second derivative penalty $\lambda_{\text{der}}$, and the main regularization parameter $\lambda$. In particular, for each $\alpha$ and each $\lambda_{\text{der}}$ on the net, we search for the optimal $\lambda$. In order to pick the initial $\lambda$, we first find the ridge estimation $\beta$ with parameter $\rho=1$. We then compute the norm of each of the groups of functional coefficients, $\|\beta^k\|$. Note that in the second update of Theorem \ref{thm: update admm group LASSO}, the soft threshold operator would eliminate all blocks if $\lambda$ is slightly higher than the maximum of these norms. On the other hand, this update would keep all the coefficients if $\lambda$ is slightly lower than the smallest norm. Therefore, a reasonable procedure is to design a grid of $\lambda$'s between a number slightly lower than the minimum norm of the blocks and a number slightly higher than the maximum norm of these block coefficients.

\section{Estimation: GMD}\label{sec: GMD}

In this section, we derive the groupwise-majorization-descent (GMD) algorithm for solving the objective functions in Section \ref{sec: model description}. Unlike the ADMM, this algorithm is geared toward the objective function with group-wise penalty terms. Motivated by \cite{yang2015fastgglasso}, we derive the GMD algorithm under our setting. In addition, we do not force the basis functions to be orthogonal, which allows us to have more flexibility. Thus, throughout this section, we use the coordinate system based on the original basis $\cal B$ without orthogonalization.


\subsection{Algorithm}
The MFG-Elastic Net problem without the orthogonalization is 
\begin{align}\label{gelast1}
\arg\min_{\beta}\frac{1}{2}\| Y- [\tilde{X}\lo {1:n}]\trans {G} [\beta] \|_2^2 + \frac{\lambda_{\text{der}}}{2}  [\beta'']^T {G}[\beta'']  +\lambda (1-\alpha) \tsum\lo {j=1}\hi p \| \beta\hi j\|\lo {\H\hi j} + \alpha \lambda \tsum\lo {j=1}\hi p \| \beta\hi j\|^2\lo {\H\hi j},
\end{align}
where the coordinates are associated with the original basis $\cal B$. This optimization problem and the following derived algorithm include the steps that also solve for the MFG-Lasso $(\alpha=0)$ and the ridge regression $(\alpha=1)$. In the equation (\ref{gelast1}), we remove $n$ for computational convenience. It will be adjusted when we seek the $\lambda_\text{der}$ and $\lambda$ in the grid construction. We define the loss function as follows.
\begin{align} \label{loss function}
    L([\beta])=  \frac{1}{2}\| Y- [\tilde{X}\lo {1:n}]\trans {G} [\beta] \|_2^2 + \frac{\lambda_{\text{der}}}{2}  [\beta'']^T {G}[\beta'']. 
\end{align}
Consequently, the objective function (\ref{gelast1}) is $L([\beta]) + g(\beta)$ where $g(\beta)= \lambda (1-\alpha) \tsum\lo {j=1}\hi p  \sqrt{[\beta^j]\trans G^j [\beta^j]} + \alpha \lambda\tsum\lo {j=1}\hi p  [\beta^j]\trans G^j [\beta^j]$.

\begin{lemma}\label{lem: qm condition}
    The loss function (\ref{loss function}) satisfies the quadratic majorization (QM) condition with $H= {G} [\tilde{X}\lo {1:n}]\trans [\tilde{X}\lo {1:n}] {G} + \lambda_{\text{der}} {B''} $. In other words, for any $\beta, \beta^* \in \H$,
    \begin{align}\label{eq: QM MFG lasso}
        L([\beta]) \le L([\beta^*]) + ([\beta]-[\beta^*])\nabla L([\beta^*]) + \frac{1}{2} ([\beta]-[\beta^*])\trans H ([\beta]-[\beta^*]),
    \end{align}
    where,
\begin{align} \label{nabla of L}
     \nabla L(\beta^*|D)= G [\tilde{X}\lo {1:n}] ( [\tilde{X}\lo {1:n}]\trans G [\beta]-Y  ) + \lambda_{\text{der}} B''[\beta^*].
\end{align}
\end{lemma}
In addition to Lemma \ref{lem: qm condition}, it is straightforward to see that if $\beta\neq \beta^*$, we have the strict inequality,
\begin{align} \label{QM condition}
    L(\beta| D) < L(\beta^*|D)-
    ([\beta]-[\beta^*])^T U(\beta^*)+ \frac{1}{2} ([\beta]-[\beta^*])\trans H ([\beta]-[\beta^*]).
\end{align}
Thus, it leads to the strict descent property of the updating algorithm. Let $\beta^*$ be the current solution to the optimization problem and $\beta$ be the next update. Assume that we update the $\beta$ for each functional predictor $j=1,\ldots, p$. In other words,  $[\beta]-[\beta^*]$ has a form of $(0, \ldots, 0, [\beta^j]- [(\beta^*)^j], 0,\ldots, 0)\trans$, which leads to simplification of the objective function in the new optimization problem. 
Let $U=-\nabla L(\beta^*)$ and $U^j$ be the sub-vector of $U$ with the indices $(m(j-1)+1, \ldots, mj)$. Let $H^j$ be the $j$-th block diagonal matrix of $H$. Then, (\ref{eq: QM MFG lasso}) is 
\begin{align*}
    L([\beta]) &\le L([\beta^*]) - ([\beta^j]-[(\beta^*)^j])U^j + \frac{1}{2} ([\beta^j]-[(\beta^*)^j])\trans H^j ([\beta^j]-[(\beta^*)^j])\\
    &\le  L([\beta^*]) - ([\beta^j]-[(\beta^*)^j])U^j + \frac{1}{2}\gamma_j ([\beta^j]-[(\beta^*)^j])\trans ([\beta^j]-[(\beta^*)^j]),
\end{align*}
where $\gamma_j$ is a value slightly larger than the largest eigenvalue of $H^j$, which further relaxes the upper bound. In practice, we take $\gamma_j=(1+\epsilon^*)\eta_j$ with $\epsilon^*=10^{-6}$ where $\eta_j$ is the largest eigenvalue of $H^j$. Finally, the update rule for $\beta^j$ is the solution to the following optimization problem.
\begin{align}\label{eq: gmd optimization 1}
    \arg\min_{\beta^j\in\H^j} -([\beta^j]-[(\beta^*)^j])U^j + \frac{1}{2}\gamma_j ([\beta^j]-[(\beta^*)^j])\trans ([\beta^j]-[(\beta^*)^j]) + g^j(\beta),
\end{align}
where $g^j$ is the $j$-th term of $g(\cdot)$. We have a closed-form solution to this problem using a similar trick of Lemma \ref{rem2}.
\begin{align}\label{eq: GMD update algorithm}
    [\beta^j]^{\text{(new)}}= \frac{1}{2\alpha \lambda + \gamma_j} S^{\H^j}_{\lambda(1-\alpha)}( U^j+ \gamma_j [\beta ^j]^{(\text{old})} ), \quad j=1,\ldots,p,
\end{align}
   where $U^j=-\nabla L([\beta ^j]^{(\text{old})})$ and 
   $\nabla L(\beta)= G [\tilde{X}\lo {1:n}] ( [\tilde{X}\lo {1:n}]\trans G[\beta] -  Y ) + {\lambda_{\text{der}}} {B''}[\beta].$

\subsection{Tuning parameter selection}
While iterating over this GMD update rule, we can reduce the computational burden more efficiently during  the tuning parameter selection with the strong rule technique. See \cite{strongrule}.

\textit{Step 1. (Initialization)}
Given $\alpha \in (0,1)$, the largest $\lambda$ in the grid points is the smallest value of $\lambda$ such that all its associated coefficients are zero. In particular, using the KKT condition (see Lemma \ref{rem2}), the largest $\lambda$ in the grid points is 
\begin{align*}
    \lambda^{(1)}=\max_j \frac{\|U^j(0)\|}{1-\alpha}.
\end{align*}
Therefore, the initial $\beta$ is zero. Then, the smallest $\lambda$ of the grid points is set to be a certain small number to include all the functional predictors, usually a fraction of the largest $\lambda$ value of the grid. 
 The process of searching for the optimal $\lambda$ starts with the largest value of the grid points and moves backward to the smallest value.
 
\textit{Step 2. (Iteration)}
At $\lambda^{(k)}$, we add $j$-th functional predictor to the active set if it satisfies the strong rule condition,
\begin{align*}
    \|U^j([\beta^j(\lambda^{(k)})])\| > (2\lambda^{(k+1)}-\lambda^{(k)})(1-\alpha),
\end{align*}
for $j=1,\ldots, p$. Subsequently, we update $\beta$ with these reduced predictors by iterating the update rule (\ref{eq: GMD update algorithm}) until numerical convergence. The stopping criteria for this iterative process can be chosen the absolute or relative. Next, in order to make sure that the strong rule does not leave out some of the worthy coefficients, we check the KKT condition on the rest of the blocks of the current solution,
\begin{align*}
    \|U^j([\beta^j_{update}(\lambda^{(k+1)})])\|<\lambda^{(k+1)}(1-\alpha),
\end{align*}
where $\beta^j_{update}(\lambda^{(k+1)})$ is the updated $\beta^j$  when the iterative GMD algorithm hits the stopping criteria on the result of the strong rule screening.
If $j$-th functional coefficient violates the KKT condition, we add it to the active set and update $\beta$ using (\ref{eq: GMD update algorithm}). This process of checking the KKT condition and updating, continues until there is no functional coefficient that violates the KKT condition. We store the solution of the final updated value to  $\beta^j(\lambda^{(k+1)})$. We use $\beta^j(\lambda^{(k+1)})$ to repeat \textit{(Step 2)} for the next value of $\lambda$ (warm start). 

It is worth mentioning that the strong rule does not allow that the main regularization for $\lambda$ to be computed in parallel because of the warm start, i.e. we search for $\lambda$ sequentially. However, the main computational cost is paid in this regularization. The strong rule allows the algorithm to enjoy predictor screening, which leads to a cost-effective computation by storing and computing on smaller size vectors. On the other hand, the strong rule does not seem to be valid for the main regularization of the ADMM algorithm because there are two objective functions involved in this algorithm. Hence, it is possible to tune the regularization parameters in parallel via ADMM.

\section{Asymptotic results}\label{sec: asymptotic}
In this section, we derive the consistency of the multivariate functional group LASSO (MFG-LASSO) when functions are fully observed. In particular, the consistency breaks down to the selection consistency and the estimation consistency, which is known as the oracle property.

We first illustrate the consistency of the operators used in the estimation procedure. Since the implementation in Section \ref{sec: coordinate representation} is based on the method of moments estimate, the following lemma is an immediate result from the functional-version of the central limit theorem in a separable Hilbert space. See \cite{hsing2015theoretical}.
\begin{lemma}\label{lem: consistency of operators}
    If $E\|X\|\hi 4 \lo {\H} <\infty$ and $EY\hi 4 < \infty$, then
    \begin{enumerate}
        \item $\sqrt{n}(\hat{\Gamma}\lo {XX} - \Gamma\lo {XX}) \cid N(0, \Sigma\lo {XX})$,
        \item $\sqrt{n}(\hat{\Gamma}\lo {YX} - \Gamma\lo {YX}) \cid N(0, \Sigma\lo {YX})$,
        \item $\sqrt{n}(\hat{\Gamma}\lo {YY} - \Gamma\lo {YY}) \cid N(0, \Sigma\lo {YY})$,
    \end{enumerate}
    where $\Sigma\lo {XX} = E[\{(X-EX)\otimes (X-EX) - \Gamma\lo {XX}\}\otimes \{(X-EX)\otimes (X-EX) - \Gamma\lo {XX}\}]$ and $\Sigma \lo {YX}$, $\Sigma \lo {YY}$ are similarly defined. 
\end{lemma}

Now, we limit our index to $\sten A$, the true active set of the population functional coefficient $\beta$. For convenience, we use the notation for truncated-version by the superscript $J$ such that $\beta \hi J = (\beta ^ {j}:j \in \sten A)\in \H \hi J$ and use $J$ and $\sten A$ interchangeably. 
\begin{lemma} \label{lem: consistency when J is known}
In addition to the assumptions in Lemma \ref{lem: consistency of operators}, assume that for any $j$, there exists ${g^j} \in H^j$ such that ${\beta}^j = \Gamma^{1/2}_{X^jX^j}({g}^j)$. This means each ${\beta}^j$ is in the range of $\Gamma_{X^j X^j}^{1/2}$. Consider $\beta^{J}_n$ as a minimizer of 
\begin{align}\label{eq: sam-objective function with known J - 1}
     \frac{1}{2}E_n[(Y-\inner{X^{J},\beta}{})^2] + \lambda_n \sum_{j \in {\sten A}} \|\beta^j \|\lo {\H^j}.
\end{align}
If $\lambda_n$ approaches zero slower than the rate at which $\sqrt{n}$ approaches infinity, $\|  \beta^{J}_n - {\beta}^{J} \|\lo{\H }$ converges to zero in probability, slightly slower than $\sqrt{\lambda_n} + \lambda^{-1}_n n^{-1/2}$.
\end{lemma}
The above lemma illustrates that if we know the true functional predictors, the solution to the optimization problem (\ref{eq: sam model objective-1}) achieves the consistency. Let $M_n(\cdot)$ be the objective function in (\ref{eq: sam-objective function with known J - 1}). Then,
\begin{align}\label{eq: sam-objective function with known J - 2}
    M_n(\beta)=\frac{1}{2}  { \hat{\Gamma}_{YY}} -    \hat{\Gamma}_{YX^{J}}\beta + \frac{1}{2} \inner{\beta,   \hat{\Gamma}_{X^{J} X^{J}} \beta}{} + \lambda_n \sum_{j \in J} \|\beta^j \|\lo {\H}.
\end{align}
Note that (\ref{eq: sam-objective function with known J - 2}) is asymptotically strictly convex as long as we can assume that $\Gamma_{X^{J} X^{J}}$ is a positive-definite operator. Similarly, the original objective function (\ref{eq: sam model objective-1}) has also a unique solution if we can assume that $\Gamma_{XX}$ exists and is positive definite. By using Lemma \ref{lem: consistency when J is known} as a bridge, we prove the consistency of our estimate in the following theorem.
\begin{theorem}\label{thm: asymptotic results}
    Assume that
    \begin{enumerate}
        \item The fourth moments of $X$ and $Y$ are bounded.
        \item For any $j$, there exists ${g^j} \in H^j$ such that ${\beta}^j = \Gamma^{1/2}_{X^jX^j}({g}^j)$. 
        \item In the population, we have such a condition that,
            \begin{align*}
            \max\limits_{i \in J^c}\| \Gamma^{1/2}_{X^i X^i} C_{X^i X^J} C_{X^J X^J}^{-1} diag((\cdot) / \| {\beta^j}\|\lo {\H}) ({g^J})  \|_{\H^J} < 1,
            \end{align*}
            where $C_{X^iX^J}$ and $C_{X^JX^J}$ are the correlation operators defined in \cite{baker}.
    \end{enumerate}
     Then, the multivariate functional group LASSO estimate satisfies the following.
     \begin{enumerate}
         \item Let $\hat{\beta}$ be the solution minimizing (\ref{eq: sam model objective-1}), and $\hat{\sten A}=\{j; \hat{\beta}\hi j \neq 0\}$ be the estimated active set. Then, $P(\hat{\sten A}=\sten A)$ converges to 1.
         \item $\|\hat{\beta}-\beta\|_{\H}\rightarrow 0$ in probability if $\lambda_n$ approaches zero slower than the rate of $n^{-1/2}$.
     \end{enumerate}
\end{theorem}
The assumption 1 is commonly used in the condition for the functional central limit theorem. The assumption 2 states that the functional coefficients $\beta$ lies in the support of the functional predictor $X$, which means that we restrict the potential range of $\beta$ to be in the range of $\Sigma_{XX}$. The assumption 3 is a modified version of the necessary condition for the LASSO to be consistent that is derived in  \cite{zou2006adaptive}.

\section{Simulation Studies} \label{sec: simulation}
In this section, we investigate the performance of the proposed method for scalar on functional penalized regressions through a simulation study.
Consider $T=[0,1]$ with a hundred observed time points equally-spaced, $\{ t_1, \dots, t_{100} \}$.  Suppose that there are $p=19$ random functional covariates, $X^j$, for $j=1, \dots , 19$, observed on a hundred time points equally-spaced in $T=[0,1]$, say $\{ t_1, \dots, t_{100} \}$. For $i=1, \ldots, n$, we first generate $X_i=(X_i^1, \dots, X_i^p)$ on 500 time points, $\{ t^*_1, \dots, t^*_{500} \}$, where $X_i\hi j$ is from a form of the Brownian motion,
$$ X^j(t_i^*) =\sum\limits^{i}_{k=1} N^j_k, $$ 
where $1 \leq k \leq 500$, $N^j_k \sim N(0,1)$. We generate the response values following the model
$$ Y= \inner{X^1  , \beta^1}{} + \inner{X^2,\beta^2}{}  + \inner{X^3, \beta^3}{} + \sigma \epsilon, $$
where $\epsilon \sim N(0,1)$, $\beta^1(t)=\sin(\frac {3 \pi t}{2})$, $\beta^2(t)=\sin(\frac{5 \pi t}{2})$, and $\beta^3(t)=t^2$ that are elements of $\H^j$ for $j=1,2,3$. Therefore, there are three functional predictors out of $19$ in the population active set, $\sten A =\{1,2,3\}$. We drop $400$ observed time points so that the remaining $100$ time points are equally spaced over $[0,1]$.

To investigate the method thoroughly, we applied different numbers of observations ($100$, $200$, $500$) and different standard deviations for the residual term $\sigma=0.01, 0.1, 1$. In each sample, we divide the observations into two sets for training and test sets ($80$\% for the training set, and $20$\% for the test set). We measure the root mean squared error (RMSE) of the prediction for the response values of the test set. In addition, we measure the number of functional predictors that are chosen correctly. More specifically, we count the correctly identified functional predictors in the population active set, the size of which is 3, and in the population inactive set, the size of which is 16  while predicting the test response values. We use $m=21$ B-spline basis functions to convert the observed values to functional objects and coordinate representations. We use 5-fold cross validation to tune the regularization parameters on a net.

In each scenario, we generate 100 samples and compute the percentages of correctly selected functional predictors that are tabulated in Table \ref{tab: selection}, and compute the mean and standard deviation of the test RMSE that are in Table \ref{tab: estimation}. Furthermore, we compare the sparse methods along with the scalar on functional ordinary least square method (OLS), ridge regression, and the oracle procedure in which only the functional predictors in the population active set are used in the OLS. For the sparse models, we apply multivariate functional group LASSO (MFG-LASSO), and the MFG-Elastic Net (MFG-EN). The two algorithms, GMD with the strong rule and ADMM, provide similar results while the GMD algorithm is much faster on serial systems and ADMM is faster on parallel computational systems. Thus, we show the results using the GMD and strong rule algorithm in this paper. 

From Table \ref{tab: selection}, we can see that the MFG-sparse methods effectively select the correct functional predictors. It also shows the consistency in an empirical way. In particular, they always select the active set correctly even with a large noise, but the selection performances of eliminating the inactive set predictors are poor with a small sample or large noise. The MFG-EN tends to choose more functional predictors than others. It is an expected result since the MFG-EN penalty includes the quadratic term which gives more stability but tends to choose more predictors.

\begin{table}[]
    \centering
    \caption{Percentages of correct selection in the test set under various simulation scenarios. In each case, 100 random samples are used. In each sample, we count the correctly identified functional predictors for the active set of the size 3 and the inactive set of the size 16. Then, we compute the overall percentage out of 100 samples.}
    \label{tab: selection}
    \begin{tabular}{|ccc|ccccc|}
    \hline
        \multicolumn{2}{|c}{Parameters}  & \multirow{2}{*}{Selection} & \multicolumn{5}{c|}{Methods}\\
        $\sigma$  &  $n$ &  & OLS & Ridge & MFG-LASSO & MFG-EN &  Oracle\\
        \hline
        \multirow{6}{*}{0.01}  &  \multirow{2}{*}{100} &   Inactive & 0  &  0  &  76  &  66  & 100 \\ 
&   &  Active & 100  &  100  &  100  &  100  & 100 \\ 
  &  \multirow{2}{*}{200} &   Inactive & 0  &  0  &  93  &  88  & 100 \\ 
&   &  Active & 100  &  100  &  100  &  100  & 100 \\ 
  &  \multirow{2}{*}{500} &   Inactive & 0  &  0  &  100  &  99  & 100 \\ 
&   &  Active & 100  &  100  &  100  &  100  & 100 \\ 
\hline
\multirow{6}{*}{0.1}  &  \multirow{2}{*}{100} &   Inactive & 0  &  0  &  73  &  64  & 100 \\ 
&   &  Active & 100  &  100  &  100  &  100  & 100 \\ 
  &  \multirow{2}{*}{200} &   Inactive & 0  &  0  &  92  &  86  & 100 \\ 
&   &  Active & 100  &  100  &  100  &  100  & 100 \\ 
 &  \multirow{2}{*}{500} &   Inactive & 0  &  0  &  100  &  99  & 100 \\ 
&   &  Active & 100  &  100  &  100  &  100  & 100 \\ 
\hline
\multirow{6}{*}{1}   &  \multirow{2}{*}{100} &   Inactive & 0  &  0  &  25  &  21  & 100 \\ 
&   &  Active & 100  &  100  &  100  &  100  & 100 \\ 
  &  \multirow{2}{*}{200} &   Inactive & 0  &  0  &  29  &  24  & 100 \\ 
&   &  Active & 100  &  100  &  100  &  100  & 100 \\ 
  &  \multirow{2}{*}{500} &   Inactive & 0  &  0  &  51  &  44  & 100 \\ 
&   &  Active & 100  &  100  &  100  &  100  & 100 \\
\hline
    \end{tabular}
\end{table}

Table \ref{tab: estimation} illustrates the estimation performance using the test RMSE. The overall behavior of the methods in terms of prediction errors is similar to that of the selection performance. As the sample size grows, the RMSEs are closer to that of the oracle estimator and their standard deviations decrease. Compared to the OLS, the sparse methods outperform when there are not enough observations or the functions are noisy. The OLS performs slightly better than the sparse methods when we have large enough $n$ and small noises. However, the standard errors of the OLS RMSE are larger than that of the MFG-methods. The ridge method is worse than the OLS with the small noise, but it is better than the OLS with the large noise. Overall, the sparse methods, MFG-LASSO and MFG-EN, perform the best in general because their results are very close to the oracle estimations. Considering that the sparse methods use much less functional predictors, the simulation results illustrate a great effectiveness of our methods in reducing both the model complexity and the prediction error.

\begin{table}[]
    \centering
    \caption{Average test RMSE of different methods under different simulation scenarios. In each case, 100 random samples are used to compute the mean and standard deviation with parentheses.}
    \label{tab: estimation}
    \begin{tabular}{|cc|ccccc|}
    \hline
        \multicolumn{2}{|c|}{Parameters}  & \multicolumn{5}{c|}{Methods}\\
        \hline
        $\sigma$  &  $n$ &  OLS & Ridge & MFG-LASSO & MFG-EN &  Oracle\\
        \hline
\multirow{6}{*}{0.01}   &  \multirow{2}{*}{100} & 1.57  &  2.41  &  1.01  &  1.02  & 0.9 \\ 
&   &  \small{(0.47)} &  \small{(0.54)} &  \small{(0.55)} &  \small{(0.55)} & \small{(0.61)} \\ 
  &  \multirow{2}{*}{200} & 0.7  &  1.22  &  0.75  &  0.76  & 0.66 \\ 
&   &  \small{(0.45)} &  \small{(0.35)} &  \small{(0.43)} &  \small{(0.43)} & \small{(0.47)} \\ 
  &  \multirow{2}{*}{500} & 0.48  &  0.72  &  0.56  &  0.57  & 0.47 \\ 
&   &  \small{(0.3)} &  \small{(0.22)} &  \small{(0.26)} &  \small{(0.26)} & \small{(0.31)} \\ 
\hline
\multirow{6}{*}{0.1}  &  \multirow{2}{*}{100} & 1.6  &  2.41  &  1.02  &  1.03  & 0.91 \\ 
&   &  \small{(0.47)} &  \small{(0.55)} &  \small{(0.55)} &  \small{(0.54)} & \small{(0.6)} \\ 
 &  \multirow{2}{*}{200} & 0.73  &  1.22  &  0.76  &  0.77  & 0.67 \\ 
&   &  \small{(0.44)} &  \small{(0.35)} &  \small{(0.43)} &  \small{(0.42)} & \small{(0.47)} \\ 
  &  \multirow{2}{*}{500} & 0.5  &  0.73  &  0.58  &  0.58  & 0.49 \\ 
&   &  \small{(0.29)} &  \small{(0.22)} &  \small{(0.26)} &  \small{(0.26)} & \small{(0.3)} \\ 
\hline
\multirow{6}{*}{1} &  \multirow{2}{*}{100} & 2.99  &  2.82  &  1.64  &  1.67  & 1.5 \\ 
&   &  \small{(0.65)} &  \small{(0.57)} &  \small{(0.44)} &  \small{(0.45)} & \small{(0.44)} \\ 
 &  \multirow{2}{*}{200} & 1.95  &  1.8  &  1.37  &  1.38  & 1.32 \\ 
&   &  \small{(0.32)} &  \small{(0.31)} &  \small{(0.31)} &  \small{(0.31)} & \small{(0.31)} \\ 
  &  \multirow{2}{*}{500} & 1.38  &  1.37  &  1.21  &  1.21  & 1.18 \\ 
&   &  \small{(0.19)} &  \small{(0.17)} &  \small{(0.17)} &  \small{(0.17)} & \small{(0.18)} \\  
\hline
    \end{tabular}
\end{table}

\begin{figure}[]
    \centering
    \includegraphics[width=1\textwidth]{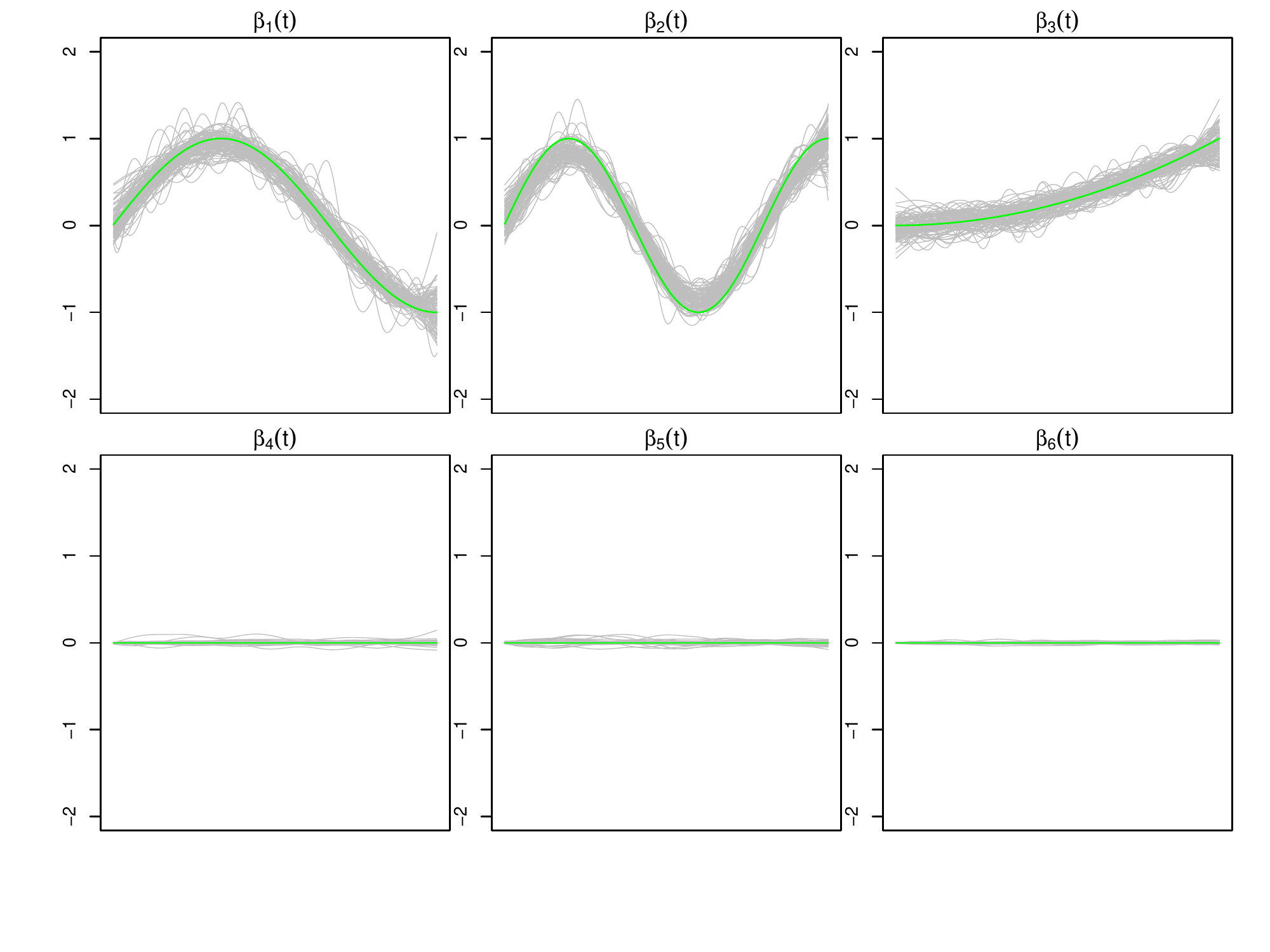} 
    \caption{This figure displays the estimated functional coefficients by the MFG-LASSO from a hundred simulated data sets when $n=100$, $\sigma=1$. The green curves are the true coefficient curves and the grey curves are the estimated coefficients. The estimated curves for the remaining of the coefficients from the seventh to the nineteenth are very similar to the fourth, fifth and sixth functions (inactive coefficients) displayed in this figure. }
    \label{fig:simul_worst}
\end{figure}

\begin{figure}[]
    \centering
    \includegraphics[width=1\textwidth]{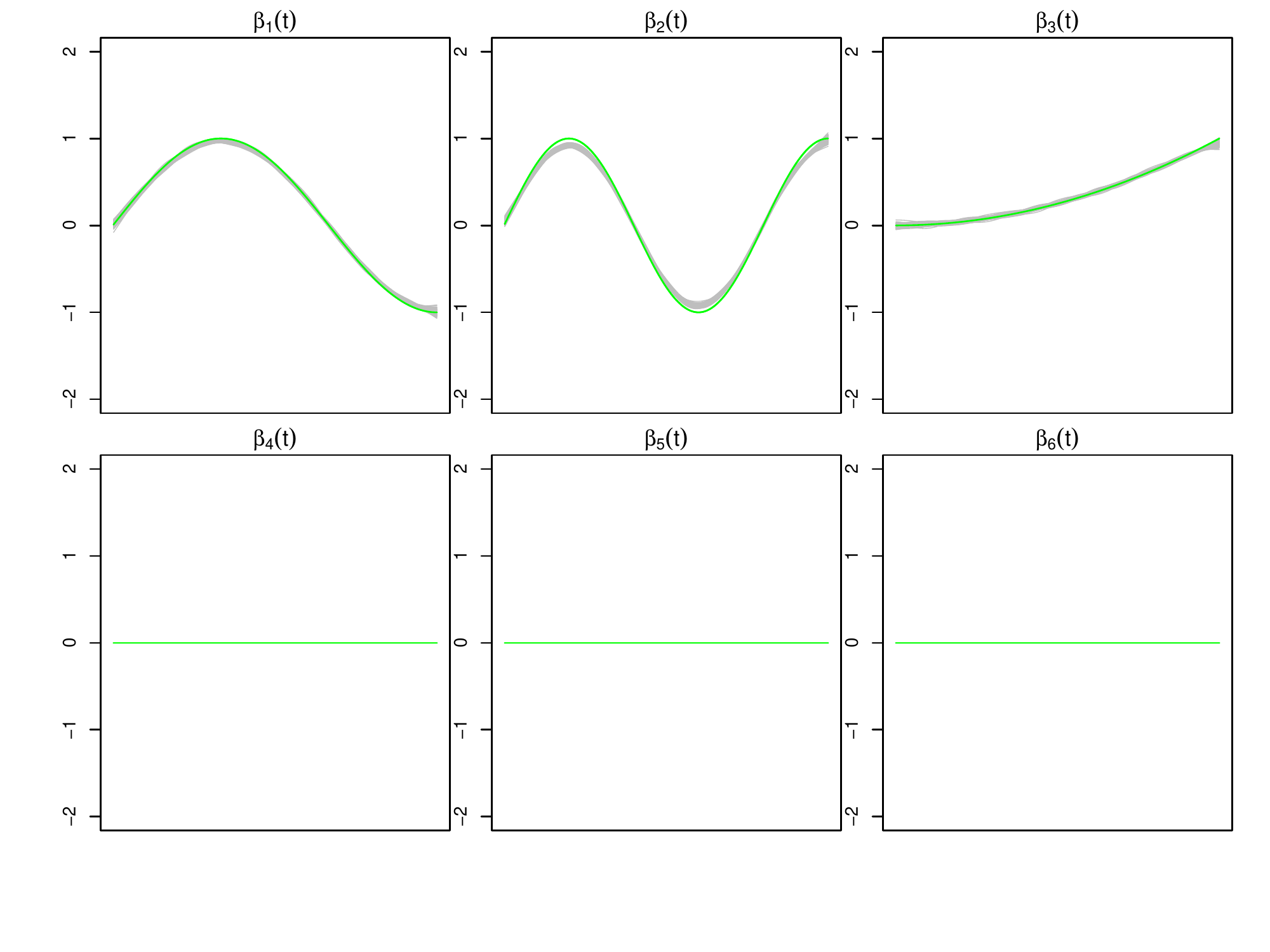} 
    \caption{This figure displays the estimated functional coefficients by the MFG-LASSO from a hundred simulated data sets when $n=500$, $\sigma=0.01$. The green curves are the true coefficient curves and the grey curves are the estimated coefficients. The estimated curves for the remaining of the coefficients from the seventh to the nineteenth are very similar to the fourth, fifth and sixth functions (inactive coefficients) displayed in this figure. }
    \label{fig:simul_best}
\end{figure}

Figure \ref{fig:simul_worst} shows the estimated functional coefficients $\hat{\beta}^1(\cdot),\ldots, \hat{\beta}^6(\cdot)$ from the MFG-LASSO in a hundred simulation samples when $n=100$, $\sigma=1$, the worst performance case. The green curves are the true functions, and the rest of the curves are the estimations. Figure \ref{fig:simul_best} shows the results when $n=500$, $\sigma=0.01$, the best performance case. 

\section{Application  to fMRI}\label{sec: real data application}
We apply our methods to a human brain fMRI data set collected by the New York University.
This data set is part of the \href{http://fcon_1000.projects.nitrc.org/indi/adhd200/}{ADHD-200} resting-state fMRI and anatomical datasets. The parent project is \href{https://www.nitrc.org/projects/fcon_1000/}{1000 Functional Connectomes Project}.
The BOLD-contrast activities of the brain are measured by the fMRI machine during a 430 seconds period of time. In order to extract the time courses, 172 equally-spaced signal values were recorded as the observed points within the 430 seconds period of time. Prior to the analysis, the automated anatomical labeling (AAL) \cite{aal} was applied to the raw fMRI data by averaging the BOLD activities of the clusters of voxels in $p=116$ regions of the brain, the regions of interest (ROI). This procedure is called masking, clustering the voxels by regions and averaging the time series signals within the region.
The data consists of between five to seven brain resting state fMRI records taken from 290 human subjects. We randomly choose two brain images from each human subject, and clean the data by removing missing response values. We choose different response values in each regression analysis, such as the subjects' intelligence quotient (IQ) scores, verbal IQ, performance IQ, attention deficit hyperactivity disorder (ADHD) index, ADHD Inattentive, and ADHD Hyper/Impulsive. Then, we split the data by 80\% for the training set and 20\% for the test set. We use $m=31$ B-spline basis functions in the function approximation procedure.

Table \ref{fmri table} describes the test RMSE and the sparsity of the regression models. The results show that the scalar on function OLS does not work in that the RMSE is higher than the standard deviation of the response values in the test set. The ridge regression has a significantly lower RMSE while it does not select functional covariates. The MFG-LASSO eliminates more than a half of the brain regions except for the performance IQ, while its RMSE is slightly higher than the MFG-EN in most cases. In terms of the RMSE, the MFG-EN performs the best while it selects more functional predictors than the MFG-LASSO. 
It is worth mentioning that when we change the proportion of the train and test data set to 90\% and 10\%, the ratio $\frac{RMSE}{\hat{\sigma}_{\text{Y}_{\text{test}}}}$ decreases significantly for sparse regressions; however, in order to be consistent with the simulations we keep the $80\%$ to $20\%$ proportions for the train and test sets.

\begin{table} 
\centering
\begin{tabular}{ |c|c|c|c|c|} 
 \hline
 Response value & Method & RMSE & Zero curves of $116$ ROI\\
 \hline
      Y=IQ score   & Least square & 19.01 & 0 \\
 Range: $73-142$    & Ridge & 5.98 &0 \\
 $\hat{\sigma}_{\text{Y}_{\text{test}}}= 13.45$ & MFG-LASSO & 6.32 &63\\
                 & MFG-EN &5.91& 10\\
  
  \hline
         Y= Verbal IQ   & Least square &23.03  & 0 \\
      Range: $65-143$  & Ridge & 7.02 &0 \\
 $\hat{\sigma}_{\text{Y}_{\text{test}}}=13.25$   & MFG-LASSO & 6.98 &68\\
  & MFG-EN &6.44& 16\\

  \hline
       Y= Performance IQ   & Least square & 19.69 & 0 \\
    Range: $72-137$  & Ridge &6.27 &0 \\
$\hat{\sigma}_{\text{Y}_{\text{test}}}=13.89$  & MFG-LASSO & 6.79 &40\\
& MFG-EN &6.06 & 10\\

  \hline
  
 Y=ADHD Index   & Least square & 28.86 & 0 \\
      Range: $40-99$  & Ridge & 8.18 &0 \\
$\hat{\sigma}_{\text{Y}_{\text{test}}}=15.22$     & MFG-LASSO & 8.49 &75\\
 & MFG-EN &8.06&  25\\
  \hline
  
                      Y=ADHD Inattentive      & Least square &  27.81 & 0 \\
    Range: $40-90$     & Ridge & 8.40 &0 \\
      $\hat{\sigma}_{\text{Y}_{\text{test}}}=15.30$          & MFG-LASSO & 9.21 & 75\\
 & MFG-EN &8.67& 30\\

  \hline
  
             Y=ADHD Hyper/Impulsive        & Least square & 26.47 & 0 \\
          Range: $41-90$  & Ridge & 7.66 &0 \\
           $\hat{\sigma}_{\text{Y}_{\text{test}}}=14.66$       & MFG-LASSO & 8.42 &60\\
 & MFG-EN &8.54& 52\\
  \hline

\end{tabular}
\caption{ The results of applying the proposed methods to the fMRI data when predicting the IQ and ADHD scores.}
\label{fmri table}
\end{table}

At the time of writing, there is no research study that uses the exact same data. However, there are articles that predict the IQ score based on human brain measurements. \cite{IQ} predicted IQ score based on structural magnetic resonance imaging (MRI). In order to predict the IQ score, they use two methods: Principal  component analysis on gray matter volume of each voxel, and Atlas-based grey matter volume while adjusting for the brain size in both methods.  The reported RMSE with  $90\%$ to $10\%$ train to test proportions in this study is $13.07$ at its best, while the standard deviation of the IQ scores in the whole sample including test set is  $\hat{\sigma}_{Y}=12.94$. Nevertheless, the MFG-LASSO provides an RMSE of $6.32$ and the MFG-EN provides $5.91$. In addition, to the higher accuracy, our methods have much less complexity of the model. \cite{IQ} selects more than $20,000$ principle features among all of the features associated with  $556,694$ voxels in the data. Meanwhile, our methods use $53$ functional predictors for MFG-LASSO and $106$ functional predictors for MFG-EN. In each functional predictor, we use 172 time points in the raw data, and we use $m=31$ basis functions in the function approximation procedure. Therefore, the proposed methods have obvious advantages in reducing the model complexity as well as achieving higher accuracy. 
Running one regression analysis with the proposed methods using the GMD\slash Strong Rule is on average around two to three minutes on a dual Core-i7 CPU with $16$ GB memory, while the mentioned article claims an equivalent computation of  $36,000$ hours using two CPU kernels and $5$ GB RAM. In addition, there is another research study, \cite{IQpred}. In this article, the RMSE does not get any better than around $14$ while data is from a combination of resting state and task fMRI, and the sparse method uses voxels' functional connectivities (Pearson correlation between BOLD time series signals) as the input features.

\begin{figure}
    \centering
     \includegraphics[width=\textwidth,height=\textheight,keepaspectratio]{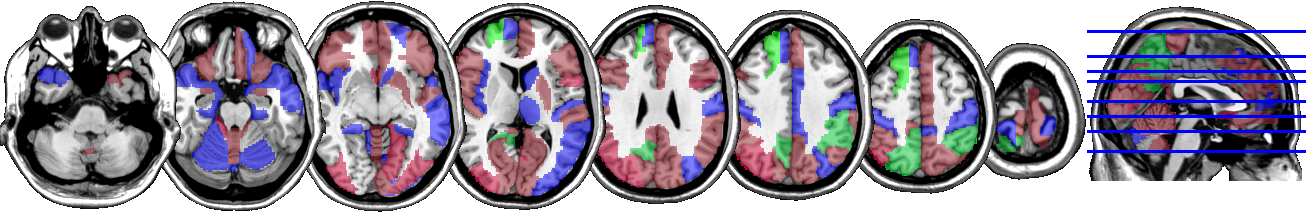}
        \includegraphics[width=\textwidth,height=\textheight,keepaspectratio]{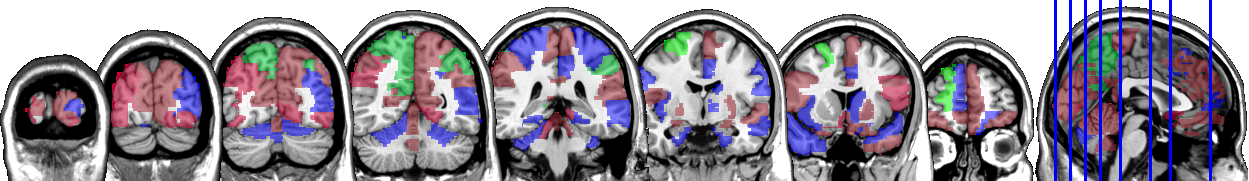}
               \includegraphics[width=\textwidth,height=\textheight,keepaspectratio]{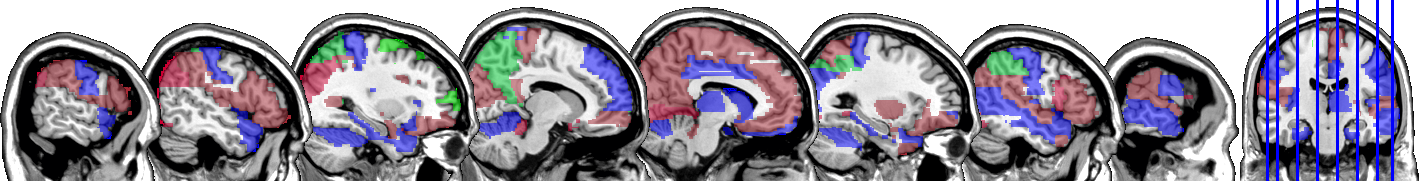}

    \caption{ The multi-slice display (Axial, Coronal, Sagittal) of the regions of interests, the BOLD activities of which achieves the lowest prediction error and correlate the most with the IQ score variability in the sample when the MFG-LASSO is used. The regions associated with the IQ score are colored red, those associated with the performance IQ are blue, and the ones  associated with the verbal IQ are colored green.}
    \label{region IQ}
\end{figure}

\begin{figure} 
    \centering
        \includegraphics[width=\textwidth,height=\textheight,keepaspectratio]{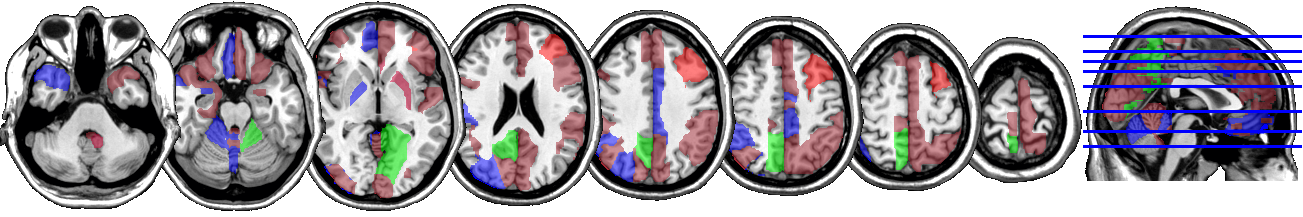}
              \includegraphics[width=\textwidth,height=\textheight,keepaspectratio]{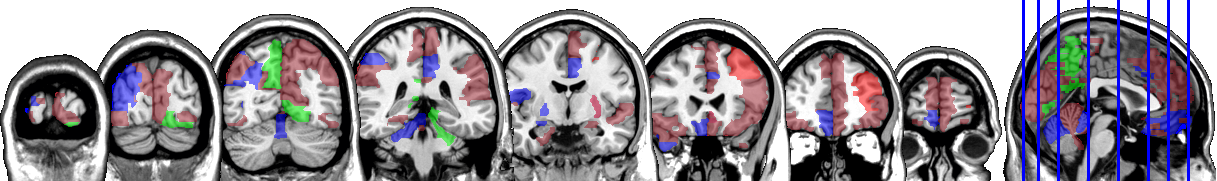}
                  \includegraphics[width=\textwidth,height=\textheight,keepaspectratio]{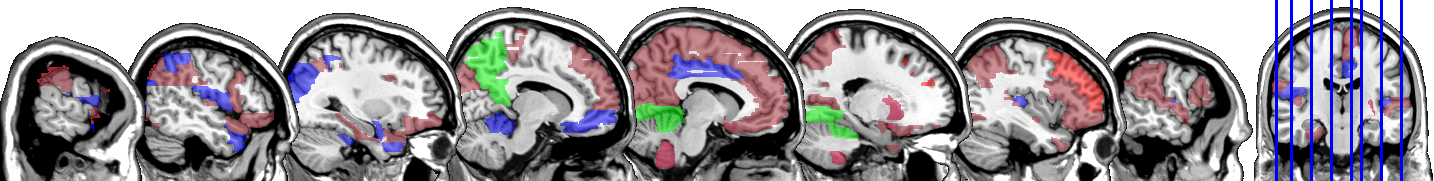}
    \caption{The multi-slice display (Axial, Coronal, Sagittal) of the regions of interests, the BOLD activities of which achieves the lowest prediction error and correlate the most with the ADHD score variability in the sample when the  MFG-LASSO is used. The regions associated with the ADHD score are colored red, those associated with the  ADHD Hyper/Impulsive are blue, and the ones  associated with the ADHD Inattentive score  are colored green.}
  \label{region adhd}
\end{figure}

In Figure \ref{region IQ} and Figure \ref{region adhd}, we  display the regions associated with the estimated active sets for IQ and ADHD by the MFG-LASSO respectively. The final active sets of the algorithms were extracted, and matched with the AAL's atlas where each of the regions has a label. The regions were manually entered into the WFU picked atlas \cite{WFU} tool of the \href{http://www.fil.ion.ucl.ac.uk/spm/} {SPM-12 } ran on MATLAB 2020b to produce mask.nii files. The mask files were imported on \href{https://people.cas.sc.edu/rorden/mricron/index.html} {MRIcron}  software to produce the multi-slice images.

The active sets cover the regions associated with IQ in \cite{nimhIQ} such as cerebello-parietal component and the frontal component. It is mentioned in the paper that the parietal and the frontal regions are strongly associated with intelligence by maintaining a connection with the cerebellum and the temporal regions. The shaded areas cover the ones mentioned in \cite{roi} as well. We provide the name of the regions associated with these active sets in the appendix. 

It is interesting that ADHD and IQ have a large proportion of common active sets. For instance, when MFG-LASSO is applied, they overlap in $35$ ROIs where the size of active sets are $53$ and $41$ for IQ and ADHD respectively. On the other hand, the ROIs that are associated with ADHD but not with IQ are the middle and superior frontal, the Parahippocampal, the inferior parietal, and the superior temporal pole gyri. The ratio of the number of right hemisphere regions to the left ones associated with IQ  is significantly greater than that of ADHD.

\section{Conclusion}\label{sec: discussion}
We propose new methods for scalar-on-function regression with the functional predictor selection along with the estimation of smooth coefficient functions when the predictors are multivariate functional data. We derive the algorithm for the implementation and develop the consistency of the methods by showing its oracle property. The simulation and real data application show the effectiveness of the methods with the superior performance of the proposed penalized methods over the functional regression model with the OLS. Furthermore, the proposed methods provide higher accuracy as well as the low complexity of the model in the fMRI study. It shows that there is an urgent need in the fields of medical sciences and other related areas. 

The manuscript also has a potential impact on the field of statistical research. Considering that there is not enough investigation made to sparse modeling of multivariate functional data, the computation algorithm derived in this paper will pave the way to develop other novel sparse methods. In addition, the methods can be extended to the nonlinear regression model via the reproducing kernel Hilbert space (RKHS). Since the theoretical justification is constructed under the infinite-dimensional setting, the extension on the RKHS can easily adopt the results from this paper. Furthermore, the proposed methods are based on groups such that a single functional predictor forms a group. Hence, it can be easily extended to the sparse models where multiple functional predictors form a group. For example, instead of averaging out fMRI signals of voxels over the regions of the brain, we would keep the original data and apply the MFG methods with groups formed by each region's voxels activities. Then, we might figure out new foundation that has been removed in the masking procedure.

\bibliography{fcs}




\newpage

\appendix
\section*{Appendix}
\noindent\textbf{Proof of Lemma \ref{lem: coordinate representation covariance operators}}
The representation of $[\hat{\Gamma}\lo {XX}]$ can be shown by the relation between the two following equations.
\begin{align*}
    \inner{f, \hat{\Gamma}\lo {XX} g}{\H}&=[f]\lo {\cal B} \trans G [X\lo {1:n}]\lo {\cal B}  Q [\hat{\Gamma}\lo {XX}]\lo {\cal B}  [g]\lo {\cal B}=E\lo n(\inner{f,X-E\lo n X}{\H}\inner{g,X-E\lo n X}{\H}),\\
    \inner{f, \hat{\Gamma}\lo {XX} g}{\H}&=[f]\lo {\cal B} \trans [\hat{\Gamma}\lo {XX}] [g]\lo {\cal B},
\end{align*}
for any $f,g\in\H$.
The second equation can be shown as following. For any $\beta \in \H$,
\begin{align*}
    \hat{\Gamma}_{YX}\beta &= E_n\{ (Y-E_nY)\otimes (X-E_nX)\} \beta = E_n\{ (Y-E_nY)\inner{X-E_nX,\beta}{\H}\}\\
    &=E_n\{ (Y-E_nY)[X-E_nX]\trans G[\beta]\}.
\end{align*}
We can also see that $\hat{\Gamma}_{XY} = n\inv [\tilde{X}_{1:n}Y]$.
\eop
\begin{lemma} \label{rem2}
Take $x,y \in \mathbb{R}^m$ where $y$ is known.
\begin{align}
    \arg\min_{x}   \left(\frac{1}{2} \|x-y\|^2  + \lambda \|x\| \right) =S_{\lambda} (y),
\end{align}
where $S_{\lambda} (y):=1_{\left \{\|y\| > \lambda \right\}}\left (1-\frac{\lambda}{\|y\|}\right )_{+}y
$ is the block soft threshold operator in real space.
\end{lemma}
\noindent\textbf{Proof of Lemma \ref{rem2}}.
Observe that
\begin{align*}
\arg\min_{x}     \left( \frac{1}{2}  (x-y)\trans (x-y)  + \lambda \|x\| \right) =\arg\min_{x} \left( \frac{1}{2} (x\trans x-2x\trans  y)  + \lambda\|x\|  \right) .
\end{align*}

To satisfy the Karush–Kuhn–Tucker (KKT) stability condition, the derivative of the above objective function with respect to $x$ must be equal to zero. If the derivative does not  exist, the subdifferential must include the zero. The derivative is $x-y+\lambda s_x$, where $s_x$ is the subdifferential of $\|x\|$ at $x$.

If $x\neq 0$, $s_x = x/\|x\|$ and the KKT condition gives 
\begin{align*}
x(1+\lambda/\|x\|)=y.
\end{align*} 
Compute the $\|y\|$ in the preceding equation and solve for $\|x\|$. Plugging it back into the equation gives us, 
$$x=(1-\lambda/\|y\|)y.$$ The condition $x\neq 0$ is equivalent to $\|y\|>\lambda$. On the other hand, $x=0$ is equivalent to  $0 \in -y+\lambda s_x$, or $y \in \lambda s_x$. In this case $s_x=\{z \in \mathbb{R}^m | \|z\| \leq 1 \} $. Therefore, $\|y\|^2 \leq \lambda^2$ which completes the proof. 
\eop

\noindent\textbf{Proof of Theorem \ref{thm: update admm group LASSO}}.\\
\indent 1) $\beta$-update.\\
Consider the objective function for $\beta$ in (\ref{eq: admm update-coordinate-scaled}). After removing the constant terms with respect to $\beta$, with the help of Lemma \ref{lem: coordinate representation covariance operators}, we have
\begin{align*}
[\beta\hi {\text{new}}]:&=\arg\min_{\beta} \left( f(\beta)+\frac{\rho}{2}([\beta]-[\gamma ]+ [U])\trans  ([\beta]-[\gamma]+ [U])  \right)\\
&=\arg\min_{\beta} \left(\frac{1}{2n}( [\beta]\trans [\tilde{X}\lo {1:n}][\tilde{X}\lo {1:n}]\trans[\beta]-2 [\beta]\trans [\tilde{X}\lo {1:n}] Y) +\frac{\rho}{2} \{ [\beta]\trans  [\beta] - 2 [\beta]\trans ([\gamma]-[U]) \} \right).
\end{align*}
Differentiate with respect to $\beta$, and set the derivative equal to zero to satisfy the KKT conditions. The result is: 
\begin{align*}
 n^{-1}[\tilde{X}\lo {1:n}] [\tilde{X}\lo {1:n}]\trans [\beta] - n^{-1}  [\tilde{X}\lo {1:n}] Y + \rho ( [\beta]- ([\gamma]-[U]))=0.
\end{align*}
Solve for $\beta$ which completes the derivation. Note that the result is similar to the functional  ridge regression. 

2) $\gamma$-update.\\
Similarly, if we remove the constant terms with respect to $\gamma$ and expand the objective function for $\gamma$, we have
\begin{align*}
    [\gamma\hi {\text{new}}]:&=\arg\min_{\gamma} \left( g(\gamma)+\frac{\rho}{2}([\beta\hi {\text{new}}]-[\gamma]+ [U])\trans  ([\beta\hi {\text{new}}]-[\gamma]+ [U])  \right)\\
    =&\arg\min_{\gamma} \left( \sum_{j=1}^p   \{ \lambda([\gamma ^j]\trans  [\gamma ^j]  )^{\frac{1}{2}}+\frac{\rho}{2} ([\gamma^j]-([(\beta^j)\hi {\text{new}}]+ [U^j]) )\trans  ([\gamma^j]-([(\beta^j)\hi {\text{new}}]+ [U^j]) ) \} \right).
\end{align*}
Note that the objective function is now additive which allows us to optimize $\gamma$ for each $\gamma ^j$, $j=1\ldots, p$. Thus, the above optimization is equivalent to
\begin{align*}
[(\gamma^j)\hi{\text{new}}]:=\arg\min_{\gamma^j} \left(\lambda([\gamma ^j]\trans  [\gamma ^j]  )^{\frac{1}{2}}+\frac{\rho}{2} ([\gamma^j]-([(\beta^j)\hi {\text{new}}]+ [U^j]) )\trans ([\gamma^j]-([(\beta^j)\hi {\text{new}}]+ [U^j]) )  \right),&
\end{align*}
for $j=1, \dots p$. Applying Lemma \ref{rem2} completes the proof.
\eop

\begin{lemma} \label{rem7}
 For $x,y \in \mathbb{R}^m$ where $y$ is known and $a,b$ are constants
\begin{align*}
    \arg\min_{x}   \left(\frac{1}{2} ( x-y)\trans (x-y) + a (x\trans
    x)^{\frac{1}{2}} + \frac{b}{2}x\trans  x \right) =\frac{1}{b+1} S_{a} (y).
\end{align*}
\end{lemma}
\noindent\textbf{Proof of Lemma \ref{rem7}}.\\
The proof is similar to that of lemma \ref{rem2}. The only difference is the derivative of the objective function. It is $x-y+a s+ bx,$ where $s$ is the subdifferential. The rest of the proof is straightforward. If $x\neq 0$ we see that
$x(1+b+\frac{a}{\|x\|})=y$. Taking norm $\|.\|$ from both sides, solving for $\|x\|$, and plugging it back, we would have $x=(\frac{1}{1+b})(1-\frac{a}{\|y \|})y$. Note that this is only possible when $\|x\| >0$, which means $\|y\| > a$. If $x=0$, it results in $0 \in -y+a s$, or $y \in a s$. Since in this case $s=\{[ Z] | Z \in \real \hi m  \& \|Z\| \leq 1 \}$, $\|y\| \leq a$, which completes the derivation above.
\eop

\noindent\textbf{Proof of Theorem \ref{thm: elastic net}}. The proof is a straightforward result from the combination of Theorem \ref{thm: update admm group LASSO} and Lemma \ref{rem7}. \eop

\begin{lemma} \label{Op}
Assume that $\Gamma_{XX}$ is a positive definite operator and when $n$ approaches infinity, $\lambda_n$ approaches zero slower than the rate at which $\sqrt{n}$ approaches infinity. Then, $\| (\hat{\Gamma}_{XX} + \lambda_n I)^{-1} \Gamma_{XX}  - ({\Gamma}_{XX} + \lambda_n I)^{-1} \Gamma_{XX}  \|_{\H} = O_p(\lambda_n^{-1}n^{-1/2})$, and $\| (\hat{\Gamma}_{XX} + \lambda_n I)^{-1} \hat{\Gamma}_{XX}  - ({\Gamma}_{XX} + \lambda_n I)^{-1} \Gamma_{XX}  \|_{\H} = O_p(\lambda_n^{-1}n^{-1/2})$, where $\|\cdot\|_\H$ is the operator norm.
\end{lemma}
\noindent\textbf{Proof of Lemma \ref{Op}}.
Note that $ \Gamma_{XX} ( \Gamma_{XX} + \lambda_n I) = I- \lambda_n( \Gamma_{XX}+ \lambda_n I)^{-1}$ and $  ( \hat{\Gamma}_{XX} + \lambda_n I)\hat{\Gamma}_{XX} = I- \lambda_n( \hat{\Gamma}_{XX}+ \lambda_n I)^{-1}$. Therefore, 
\begin{align*}
    &(\hat{\Gamma}_{XX}+ \lambda_n I)^{-1} - ({\Gamma}_{XX}+ \lambda_n I)^{-1}=(\hat{\Gamma}_{XX}+ \lambda_n I)^{-1}(\Gamma_{XX}- \hat{\Gamma}_{XX})({\Gamma}_{XX}+ \lambda_n I)^{-1}.
\end{align*}
To be specific, if we add and subtract $\lambda_n(\hat{\Gamma}_{XX}+ \lambda_n I)^{-1}(\Gamma_{XX}+ \lambda_n I)^{-1}$ in the left hand side of the above equation, we can easily derive the right hand side of the equation. In addition, we have
\begin{align} \label{nine}
    &(\hat{\Gamma}_{XX}+ \lambda_n I)^{-1} \Gamma_{XX} - ({\Gamma}_{XX}+ \lambda_n I)^{-1} \Gamma_{XX}= (\hat{\Gamma}_{XX}+ \lambda_n I)^{-1}(\Gamma_{XX}- \hat{\Gamma}_{XX})({\Gamma}_{XX}+ \lambda_n I)^{-1} \Gamma_{XX}. 
\end{align}
Note that $ (\hat{\Gamma}_{XX}+ \lambda_n I)^{-1} = (\Gamma_{XX}+O_p(n^{-1/2}) + \lambda_n I)^{-1}$ by Lemma \ref{lem: consistency of operators}. Thus, its norm is $\|(\hat{\Gamma}_{XX}+ \lambda_n I)^{-1}\|_{\H} = O_p(\lambda_n^{-1})$. By Lemma \ref{lem: consistency of operators}, $\|(\Gamma_{XX}- \hat{\Gamma}_{XX})\|_\H = O_p(n^{-1/2})$. The norm of product of the last two parentheses is bounded by $1$. Hence, $\| (\hat{\Gamma}_{XX} + \lambda_n I)^{-1} \Gamma_{XX}  - ({\Gamma}_{XX} + \lambda_n I)^{-1} \Gamma_{XX}  \|_{\H} = O_p(\lambda_n^{-1}n^{-1/2})$.\\

For the second convergence rate, note that
\begin{align*}
     (\hat{\Gamma}_{XX} + \lambda_n I)^{-1} \hat{\Gamma}_{XX}  -  (\hat{\Gamma}_{XX} + \lambda_n I)^{-1} {\Gamma}_{XX}  =(\hat{\Gamma}_{XX} + \lambda_n I)^{-1} (\hat{\Gamma}_{XX} -{\Gamma}_{XX})=  O_p(\lambda_n^{-1}n^{-1/2}).
\end{align*}
Therefore,
\begin{align*}
    \| (\hat{\Gamma}_{XX} + \lambda_n I)^{-1} \hat{\Gamma}_{XX}  - ({\Gamma}_{XX} + \lambda_n I)^{-1} \Gamma_{XX}  \|_{\H}
    &\le  \| (\hat{\Gamma}_{XX} + \lambda_n I)^{-1} \hat{\Gamma}_{XX} -   (\hat{\Gamma}_{XX} + \lambda_n I)^{-1} {\Gamma}_{XX} \|_{\H}\\
    &+ \| (\hat{\Gamma}_{XX} + \lambda_n I)^{-1} {\Gamma}_{XX}  - ({\Gamma}_{XX} + \lambda_n I)^{-1} \Gamma_{XX}  \|_{\H}\\
    &=  O_p(\lambda_n^{-1}n^{-1/2}).
\end{align*}
\eop
\noindent\textbf{Proof of Lemma \ref{lem: consistency when J is known}}.

The following proof is similar to the proof mentioned in \cite{bach} which considers a different penalty term that is square of the group LASSO penalty. Then, they proved the consistency by stating that the solution path of the group LASSO will be the same. Instead, we consider the a different optimization problem $\tilde{M}_n(.)$ proposed below, which directly leads to the consistency of multivariate functional group LASSO.

Denote $\tilde{\beta}^{J}_n$ as the unique minimizer of the following objective function.
\begin{align*}
    \tilde{M}_n(\alpha)=\frac{1}{2}  { \hat{\Gamma}_{YY}} -    \hat{\Gamma}_{YX^{J}} (\alpha) + \frac{1}{2} \inner{\alpha,   \hat{\Gamma}_{X^{J} X^{J}} (\alpha)}{} + \frac{\lambda_n}{2}
    \sum_{j\in J} \frac{\|\alpha^j \|\lo {\H^j} ^2}{\| {\beta} ^j\|\lo {\H^j}}, \quad \alpha \in \H,
\end{align*}
where $\beta^j$ is the $j$-th functional component of $\beta^J$ in the population model.
$\tilde{\beta}^{J}_n$ has a closed form solution similar to the solution of a functional predictor ridge regression $$\tilde{\beta}^{J}_n=( \hat{\Gamma}_{X^{J} X^{J}} + \lambda_n D )^{-1} ( \hat{\Gamma}_{X^{J}Y}),$$ where $D$ is a diagonal operator, $diag((\cdot)/\|{\beta}^j \|)$. We can replace $\hat{\Gamma}_{X^{J}Y}$ by the following expression, after adding and subtracting $  \hat{\Gamma}_{X^{J} X^{J}}({\beta}^{J})$.
\begin{align} \label{closed}
   \tilde{\beta}^{J}_n=( \hat{\Gamma}_{X^{J} X^{J}} + \lambda_n D )^{-1} ( \hat{\Gamma}_{X^{J} X^{J}}{\beta}^{J}+\hat{\Gamma}_{X\epsilon}),
\end{align}
where  $\hat{\Gamma}_{X\epsilon}$ is the empirical covariance operator between observed functional data $X$ and the population error, $\epsilon= Y-\inner{X,{\beta}}{}=Y-\inner{X^{J}, {\beta}^{J}}{}$. $D$ is a self-adjoint operator, and $\|{\beta}^j\|\lo {\H} \neq 0$ for all $j \in J$ by the definition of the population active set $J$. This means there are positive constants $D_{\min}= 1/{\max\limits_{j \in J}{\| {\beta}^j \|\lo {\H}
}}$ and $D_{\max}= 1/{\min\limits_{j \in J}{\| {\beta}^j \|\lo {\H}}}$ such that $D_{\max} I \succcurlyeq D  \succcurlyeq D_{\min} I$. The closed form solution (\ref{closed}) can be broken down into multiple terms. One of the term is
\begin{align}\label{sixteen}
    ( \hat{\Gamma}_{X^{J} X^{J}} + \lambda_n D )^{-1} (\hat{\Gamma}_{X\epsilon}).
\end{align}
Applying the same technique in the proof of Lemma \ref{Op} and using the result of Lemma \ref{lem: consistency of operators}, we can see that $\|  \hat{\Gamma}_{X^{J} X^{J}} + \lambda_n D^{-1} \|\lo {\H} \leq D^{-1}_{\min} \lambda_n^{-1}$, and
\begin{align*}
    ( \hat{\Gamma}_{X^{J} X^{J}} + \lambda_n D )^{-1} (\hat{\Gamma}_{X\epsilon}) = O_p(n^{-1/2}\lambda_n^{-1}).
\end{align*}
Hence, we have
\begin{align}
\begin{split}\label{eq: bnjtilde - bj}
        \tilde{\beta}^{J}_n-{{\beta}^{J}} 
    &= ( \hat{\Gamma}_{X^{J} X^{J}} + \lambda_n D )^{-1} ( \hat{\Gamma}_{X^{J} X^{J}}\beta^J+\hat{\Gamma}_{X\epsilon})- {{\beta}^{J}} \\
    &=( \hat{\Gamma}_{X^{J} X^{J}} + \lambda_n D )^{-1} ( \hat{\Gamma}_{X^{J} X^{J}}\beta^J) - ( {\Gamma}_{X^{J} X^{J}} + \lambda_n D )^{-1} {\Gamma}_{X^{J} X^{J}} \beta^J\\
    &+( {\Gamma}_{X^{J} X^{J}} + \lambda_n D )^{-1} {\Gamma}_{X^{J} X^{J}} \beta^J- {{\beta}^{J}} +  O_p(n^{-1/2}\lambda_n^{-1})
\end{split}
\end{align}
The first two terms of the last equation in (\ref{eq: bnjtilde - bj}) is $O_p(n^{-1/2}\lambda_n^{-1})$ by Lemma \ref{Op}. By using $ ( {\Gamma}_{X^{J} X^{J}} + \lambda_n D )^{-1}{\Gamma}_{X^{J} X^{J}} =I-\lambda_n ( {\Gamma}_{X^{J} X^{J}} + \lambda_n D )^{-1} D $, we can simplify the third and fourth terms of (\ref{eq: bnjtilde - bj}) as
\begin{align}\label{eq:gamma inverse gamma beta}
     ( {\Gamma}_{X^{J} X^{J}} + \lambda_n D )^{-1} {\Gamma}_{X^{J} X^{J}} \beta^J- {{\beta}^{J}} = (-\lambda_n ( {\Gamma}_{X^{J} X^{J}} + \lambda_n D )^{-1} D) \beta^J.
\end{align}
Consequently, we have 
  \begin{align}
   \tilde{\beta}^{J}_n-{{\beta}^{J}}=
   (-\lambda_n ( {\Gamma}_{X^{J} X^{J}} + \lambda_n D )^{-1} D) \beta^J + O_p(n^{-1/2}\lambda_n^{-1}).
\end{align}
Now, we show the norm of $\lambda_n ( {\Gamma}_{X^{J} X^{J}} + \lambda_n D )^{-1} D$ is $O_p(\sqrt{\lambda_n}+ n^{-1/2}\lambda_n^{-1})$. Let $h^J \in \H^J$ be the element in the assumption such that $\beta^J= {\Gamma}_{X^{J} X^{J}}^{1/2}h^J$.  Then,
\begin{align*}
    \|\lambda_n ( {\Gamma}_{X^{J} X^{J}} + \lambda_n D )^{-1} D\beta^J \|^2_{\H^J}
    &= \lambda_n^2 \langle {{{\beta}^{J}}}, D ( {\Gamma}_{X^{J} X^{J}} + \lambda_n D )^{-2}   D\beta^J\rangle_{\H^J}\\
   &\leq \lambda_n^2 D^2_{\max} \langle {{{\beta}^{J}}}, ( {\Gamma}_{X^{J} X^{J}} + \lambda_n D_{\min}I )^{-2}   \beta^J\rangle_{\H^J}\\
  &\leq \lambda_n D^2_{\max} D^{-1}_{\min} \langle {{{\beta}^{J}}},  ( {\Gamma}_{X^{J} X^{J}} + \lambda_n D_{\min}I )^{-1}   \beta^J\rangle_{\H^J}\\ 
     &=\lambda_n D^2_{\max} D^{-1}_{\min} \langle \Gamma^{1/2}_{X^{J} X^{J}} {h}^{J} , ( {\Gamma}_{X^{J} X^{J}} + \lambda_n D_{\min}I )^{-1}  \Gamma^{1/2}_{X^{J} X^{J}} {h}^{J}\rangle_{\H^J}\\
   &\leq \lambda_n D^2_{\max} D^{-1}_{\min} \|{h}^{J}\|_{\H}^2.
\end{align*}
The third line of the above equation is valid because $ \| {\Gamma}_{X^{J} X^{J}} + \lambda_n D_{\min}I\|_{\H^J} \geq \lambda_n D_{\min}$. Combining the results above, we have
\begin{align*}
    \|\tilde{\beta}^{J}_n-{{\beta}^{J}} \|\lo {\H}= O_p(\sqrt{\lambda_n}+ n^{-1/2}\lambda_n^{-1}).
\end{align*}
Now, let's compare $\tilde{\beta}^{J}_n$ and ${\beta}^{J}_n$ where ${\beta}^{J}_n$ is the solution to the optimization problem of $M_n(\alpha)$. Consider the following equation.
\begin{align}\label{diffeq}
M_n(\alpha)-\tilde{M_n}(\alpha)= \lambda_n  \sum_{j \in J} \left( \|\alpha^j \|_{\H^j}- \frac{\|\alpha^j \|_{\H^j}^2}{2 \| {\beta} ^j\|_{\H^j}} \right).
\end{align}
The partial Fr\'echet derivative of the equation (\ref{diffeq}) with respect to $\alpha^i$ for an $i\in J$ is
\begin{align}\label{thirtyfive}
D_{\alpha^i}  (M_n(\alpha)-\tilde{M_n}(\alpha))= \lambda_n  \left ( \frac{\inner{\alpha^i,\cdot}{\H^i}}{\|\alpha^i \|_{\H^i}}- \frac{\inner{\alpha^i,\cdot}{\H^i}}{\| {\beta^i} \|_{\H^i}}  \right ). 
\end{align}
Since ${\beta}^{J}$ are nonzero, (\ref{thirtyfive}) is continuously differentiable around ${\beta}^{J}$, and $D_{\alpha^i}\tilde{M_n}(\tilde{\beta}^{J}_n))=0$, we have
\begin{align*}
\| D_{\alpha^i}{M_n}(\tilde{\beta}^{J}_n))  -0\| = \lambda_n  \left \| \frac{\inner{\tilde{\beta}^{i}_n,\cdot}{\H^i}}{\|\tilde{\beta}^{i}_n \|_{\H^i}}- \frac{\inner{\tilde{\beta}^{i}_n,\cdot}{\H^i}}{\| {\beta^i} \|_{\H^i}}  \right \|,
\end{align*}
where the $\|\cdot\|$ is the operator norm. In addition, since $\beta^i\neq 0$ for $i\in J$, it can be easily shown that 
\begin{align*}
    \| D_{\alpha^i}{M_n}(\tilde{\beta}^{J}_n))  -0\|_{\H^i} \leq C \lambda_n \| {\beta}^{J}  - {\tilde{\beta}^{J}_n} \|_{\H^J},
\end{align*}
for some constant $C>0$. Thus, we have
\begin{align}\label{thirtyseven}
    \| D_{\alpha^i}{M_n}(\tilde{\beta}^{J}_n)) \|_{\H^i} =\lambda_n O_p(\lambda^{1/2}_n+n^{-1/2} \lambda^{-1
    }_n).
\end{align}
Now, since $M_n$ is strictly convex near the true $\beta^J$, its second-order Fr\'echet derivative has a lower bound. Consequently, we have
\begin{align*}
    {M_n}(\alpha^{J}) \geq {M_n}(\tilde{\beta}^{J}_n)+ \langle D_{\alpha^{J}}{M_n}(\tilde{\beta}^{J}_n), ({\alpha^{J}}-\tilde{\beta}^{J}_n)  \rangle_{\H^J} +C' \lambda_n \| \alpha^{J}- \tilde{\beta}_n^J\|^2_{\H^J},
\end{align*}
for some $C'>0$. Suppose that $\alpha^J$ is near $\tilde{\beta}_n^J$ and let $\eta_n= \| \alpha^{J}- \tilde{\beta}_n^J\|^2_{\H^J}$ which tends to zero. Subsequently, we can rewrite the lower bound such that
\begin{align}\label{ineqMn}
    {M_n}(\alpha^{J}) \geq {M_n}(\tilde{\beta}^{J}_n)+\eta_n \lambda_n O_p(\sqrt{\lambda}_n+n^{-1/2} \lambda^{-1
    }_n) +C' \lambda_n \eta^2_n,
\end{align}
If the last term is tending to zero slower than the second term, we can conclude that all minima of $M_n(\cdot)$ is inside the ball $\{\alpha^J:  \| \alpha^{J}- \tilde{\beta}_n^J\|^2_{\H^J} < \eta \}$ with probability tending to one. This is because $M_n(\cdot)$, on the edge of the ball, takes values greater the ones inside the ball. i.e., the global minimum of $M_n(\cdot)$ is at most $\eta_n$ away from $\tilde{\beta}_n^J$. Thus, the necessary condition for the proof is $\eta_n\lambda_n^{3/2} = o(\lambda_n\eta_n^2)$ and $ n^{-1/2}\eta_n=o(\lambda_n\eta_n^2)$. All together, we have the consistency results if $\eta_n$ converges to zero slower than $\lambda^{1/2}_n+n^{-1/2} \lambda^{-1}_n$. 
\eop

\noindent\textbf{Proof of Theorem \ref{thm: asymptotic results}}.
We rewrite the multivariate functional group LASSO objective function (\ref{eq: sam model objective-1}) as,
\begin{align*}
     \hat{M}_n(\alpha)=\frac{1}{2}  { \hat{\Gamma}_{YY}} -   \hat{\Gamma}_{YX} \alpha + \frac{1}{2} \inner{\alpha,   \hat{\Gamma}_{X X} \alpha}{\H} + \lambda_n\sum_{j=1}^p \|\alpha^j \|\lo {\H^j}.
\end{align*}
Denote a minimizer of $\hat{M}_n(\cdot)$ by $\hat{\beta}_n$. Since it is a convex function, it has a unique minimizer. In addition, if $\lambda_n$ goes to zero, the objective function converges to the regression problem without the penalty whose unique minimizer is $\beta$. Thus, it is easy to see that $\hat{J}=\{j:\hat{\beta}\hi j_n(\cdot) \neq 0\}$ converges to $J$ via the M-estimation theory. See \cite{van2000asymptotic} and \cite{knight2000asymptotics}.

Now, we extend $\beta^{J}_n$ in Lemma \ref{lem: consistency when J is known} with zero functions as $\beta^i_n$ for $i \in J^c$, name it $\beta_n \in \H$. Note that, it is a consistent estimator of $\beta$ by Lemma \ref{lem: consistency when J is known}. Since both of the  $M_n(\cdot)$ and $\hat{M}_n(\cdot)$ have unique minimizers and the $\beta_n$ is a consistent estimator of $\beta$, the consistency of $\hat{\beta}_n$ can be shown, if we can show that $\beta_n$ satisfies the optimal conditions for $\hat{M}_n(\cdot)$ with a probability tending to one. The (asymptotically) optimal conditions of $\hat{M}_n(\cdot)$ are 
\begin{equation*}\label{}
\begin{cases} 
              \| \hat{\Gamma}_{X^iX} \alpha -\hat{\Gamma}_{X^iY}  \|_{\H^i} \leq \lambda_n & i \notin J          \\
      \inner{\hat{\Gamma}_{X^jX} \alpha,\cdot}{\H^j} -\hat{\Gamma}_{YX^j}(\cdot) =- \frac{\lambda_n }{\| \alpha^j\|\lo {\H^j}}
      \inner{\alpha^j,\cdot}{\H^j} & j \in J. 
    \end{cases}
    \end{equation*}
The second equation is immediately satisfied with $\alpha=\beta_n$, since it satisfies the KKT condition for ${M}_n(\cdot)$. We focus on the above inequality of the optimal condition. The first derivative condition for minimizing $M_n(\cdot)$ implies that $\beta^J_n$ should justify the following equation
\begin{align*}
-\hat{\Gamma}_{YX^J}(\cdot)+ \inner{\hat{\Gamma}_{X^J X^J} \beta_n^J, \cdot}{\H^J}+ \lambda_n\sum_{j\in J} \frac{\inner{\beta_n^j,\cdot}{\H^j}}{\|\beta_n^j \|\lo {\H^j}}=0.
\end{align*}
Define $D_n$ be a operator from $\H^J$ to $\H^J$ such that $D_n(\alpha^J)=diag(\alpha^{j}/\| \beta_n^j\|)$ for $j \in J$.
We rewrite the above equation as
\begin{align*}
-\hat{\Gamma}_{YX^J}(\cdot)+ \inner{(\hat{\Gamma}_{X^J X^J}+\lambda_n D_n) \beta_n^J, \cdot}{\H^J}=0.
\end{align*}
In addition, note that 
\begin{align*}
    \hat{\Gamma}_{YX^J}(\cdot) = \inner{ \hat{\Gamma}_{X^JY}, \cdot}{\H^J} =  \inner{ \hat{\Gamma}_{X^JX^J}\beta^J+ \hat{\Gamma}_{X\epsilon}, \cdot}{\H^J} .
\end{align*}
Thus, we have
\begin{align*}
   \inner{\beta_n^J,\cdot}{}=\inner{(\hat{\Gamma}_{X^J X^J}+\lambda_n D_n)^{-1} ( \hat{\Gamma}_{X^{J} X^{J}}{\beta}^{J}+\hat{\Gamma}_{X^J \epsilon}),\cdot}{\H^J}.
\end{align*}
Furthermore, by using a similar technique used in (\ref{eq:gamma inverse gamma beta}), 
\begin{align*}
    (\hat{\Gamma}_{X^J X^J}+\lambda_n D_n)^{-1} \hat{\Gamma}_{X^{J} X^{J}}{\beta}^{J} =\beta^J-( \hat{\Gamma}_{X^{J} X^{J}} + \lambda_n D_n )^{-1}\lambda_n D_n \beta^J.
\end{align*}
Thus, for an $i \in J^c$:
\begin{align*}
    \hat{\Gamma}_{X^iY}-\hat{\Gamma}_{X^iX^J} \beta_n^J
    &= \hat{\Gamma}_{X^iY}-\hat{\Gamma}_{X^iX^J} \beta^J + \lambda_n\hat{\Gamma}_{X^iX^J}( \hat{\Gamma}_{X^{J} X^{J}} + \lambda_n D_n )^{-1} D_n \beta^J \\
    &- \hat{\Gamma}_{X^iX^J}( \hat{\Gamma}_{X^{J} X^{J}} + \lambda_n D_n )^{-1}\hat{\Gamma}_{X^J \epsilon}\\
    &=\lambda_n\hat{\Gamma}_{X^iX^J}( \hat{\Gamma}_{X^{J} X^{J}} + \lambda_n D_n )^{-1} D_n \beta^J  + \hat{\Gamma}_{X^i \epsilon}\\
    &- \hat{\Gamma}_{X^iX^J}( \hat{\Gamma}_{X^{J} X^{J}} + \lambda_n D_n )^{-1}\hat{\Gamma}_{X^J \epsilon},
\end{align*}
by using the fact that $ \hat{\Gamma}_{X^i Y}- \hat{\Gamma}_{X^{i} X^{J}}({\beta}^{J})=\hat{\Gamma}_{X^i \epsilon}$. At this point, the formulation has a similar form, derived in Theorem 11 of \cite{bach}. Furthermore, Lemma \ref{lem: consistency when J is known} satisfies  the condition that is necessary to derive the rest of the proof so that they can be derived in a similar way. 
\eop

\medskip

\noindent\textbf{List of regions of interests}:
The following are the lists of the regions of interest of the human brain used in the application section \ref{sec: real data application}. The atlas labels of the human brain and full names can be found at \href{journals.plos.org/plosone/article/file?type=supplementary&id=info:doi/10.1371/journal.pone.0088690.s001}{Atlas Label}.

\medskip
 \small
\noindent The list of the regions of interest associated with the active set of MFG-Lasso when the response value is IQ score:\\ 
\scriptsize
"Frontal--Mid--Orb--L", "Frontal--Mid--Orb--R", "Frontal--Inf--Oper--L",  
"Frontal--Inf--Oper--R","Frontal--Inf--Tri--L", "Frontal--Inf--Tri--R",   
"Frontal--Inf--Orb--L","Frontal--Inf--Orb--R", "Rolandic--Oper--R",     
"Supp--Motor--Area--L", "Olfactory--L",       "Olfactory--R",         
"Frontal--Sup--Medial--L","Frontal--Med--Orb--L", "Frontal--Med--Orb--R",  
"Rectus--L",          "Cingulum--Ant--L",    "Cingulum--Post--L",     
"Cingulum--Post--R",   "Amygdala--L",        "Amygdala--R",          
"Calcarine--L",       "Calcarine--R",       "Cuneus--L",            
"Cuneus--R",          "Lingual--L",        "Occipital--Sup--L",    
"Occipital--Sup--R",   "Occipital--Mid--R",   "Occipital--Inf--L",     
"Occipital--Inf--R",   "Parietal--Sup--L",    "Parietal--Inf--R",      
"SupraMarginal--L",   "SupraMarginal--R",   "Angular--L",           
"Angular--R",         "Precuneus--L",       "Paracentral--Lobule--L",
"Paracentral--Lobule--R","Putamen--L",         "Pallidum--R",          
"Heschl--R",          "Temporal--Sup--L",    "Temporal--Pole--Mid--L", 
"Cerebellum--3--L",     "Cerebellum--3--R",     "Vermis--1--2",          
"Vermis--3",         "Vermis--4--5",        "Vermis--6",            
"Vermis--9",          "Vermis--10".\\
 \small 
\\
\noindent The list of the regions of interest associated with the active set of MFG-Lasso when the response value is Verbal IQ:\\ 
\scriptsize
"Frontal--Sup--R",    "Frontal--Mid--Orb--L","Frontal--Mid--Orb--R",  
"Frontal--Inf--Oper--R","Frontal--Inf--Tri--L","Frontal--Inf--Tri--R",  
"Frontal--Inf--Orb--L","Frontal--Inf--Orb--R","Rolandic--Oper--R",    
"Supp--Motor--Area--L","Olfactory--L",      "Frontal--Sup--Medial--L"
"Frontal--Med--Orb--L","Frontal--Med--Orb--R","Rectus--L",           
"Cingulum--Ant--L",   "Cingulum--Post--L",  "Cingulum--Post--R",    
"Amygdala--L",       "Amygdala--R",       "Calcarine--L",        
"Calcarine--R",      "Cuneus--L",         "Cuneus--R",           
"Occipital--Sup--L",  "Parietal--Sup--L",   "Parietal--Sup--R",     
"Parietal--Inf--L",   "Parietal--Inf--R",   "SupraMarginal--L",    
"SupraMarginal--R",  "Angular--L",        "Precuneus--L",        
"Precuneus--R",      "Paracentral--Lobule--L","Paracentral--Lobule--R"
"Putamen--L",        "Pallidum--R",       "Heschl--R",           
"Temporal--Sup--L",   "Temporal--Pole--Mid--L","Cerebellum--3--L",      
"Vermis--1--2",       "Vermis--3",         "Vermis--4--5",         
"Vermis--6",         "Vermis--9",         "Vermis--10", .\\
\small 
\\
\noindent The list of the regions of interest associated with the active set of MFG-Lasso when the response value is Performance IQ:\\ 
\scriptsize
"Frontal--Sup--Orb--L","Frontal--Mid--Orb--L","Frontal--Mid--Orb--R",  
"Frontal--Inf--Oper--L","Frontal--Inf--Oper--R","Frontal--Inf--Tri--L",  
"Frontal--Inf--Tri--R","Frontal--Inf--Orb--L","Frontal--Inf--Orb--R",  
"Rolandic--Oper--R",  "Supp--Motor--Area--L","Olfactory--L",        
"Olfactory--R",      "Frontal--Sup--Medial--L","Frontal--Sup--Medial--R"
"Frontal--Med--Orb--L","Frontal--Med--Orb--R","Rectus--L",           
"Insula--R",         "Cingulum--Ant--L",   "Cingulum--Mid--L",     
"Cingulum--Post--L",  "Cingulum--Post--R",  "ParaHippocampal--L",  
"ParaHippocampal--R","Amygdala--L",       "Amygdala--R",         
"Calcarine--L",      "Calcarine--R",      "Cuneus--L",           
"Cuneus--R",         "Lingual--L",        "Occipital--Sup--L",    
"Occipital--Sup--R",  "Occipital--Mid--L",  "Occipital--Mid--R",    
"Occipital--Inf--L",  "Occipital--Inf--R",  "Postcentral--L",      
"Postcentral--R",    "Parietal--Sup--L",   "Parietal--Sup--R",     
"Parietal--Inf--L",   "Parietal--Inf--R",   "SupraMarginal--L",    
"SupraMarginal--R",  "Angular--L",        "Angular--R",          
"Precuneus--L",      "Precuneus--R",      "Paracentral--Lobule--L"
"Paracentral--Lobule--R","Caudate--L",        "Putamen--L",          
"Pallidum--R",       "Thalamus--L",       "Heschl--L",           
"Heschl--R",         "Temporal--Sup--L",   "Temporal--Pole--Sup--L",
"Temporal--Pole--Sup--R","Temporal--Mid--L",   "Temporal--Pole--Mid--L",
"Temporal--Pole--Mid--R","Cerebellum--3--L",    "Cerebellum--3--R",      
"Cerebellum--4--5--R",  "Cerebellum--6--L",    "Cerebellum--6--R",      
"Vermis--1--2",       "Vermis--3",         "Vermis--4--5",         
"Vermis--6",         "Vermis--7",         "Vermis--9",           
"Vermis--10".\\
\small 
\\
\noindent The list of the regions of interest associated with the active set of MFG-Lasso when the response value is ADHD score:\\ 
\scriptsize
"Frontal--Mid--L",    "Frontal--Mid--Orb--L","Frontal--Mid--Orb--R",  
"Frontal--Inf--Oper--L","Frontal--Inf--Oper--R","Frontal--Inf--Tri--L",  
"Frontal--Inf--Orb--L","Frontal--Inf--Orb--R","Supp--Motor--Area--L",  
"Olfactory--L",      "Frontal--Sup--Medial--L","Frontal--Sup--Medial--R"
"Frontal--Med--Orb--L","Rectus--L",         "Cingulum--Ant--L",     
"Cingulum--Post--L",  "ParaHippocampal--R","Amygdala--L",         
"Calcarine--L",      "Cuneus--L",         "Cuneus--R",           
"Occipital--Inf--L",  "Occipital--Inf--R",  "Parietal--Sup--L",     
"Parietal--Inf--L",   "SupraMarginal--L",  "SupraMarginal--R",    
"Angular--L",        "Angular--R",        "Precuneus--L",        
"Paracentral--Lobule--L","Paracentral--Lobule--R","Putamen--L",          
"Heschl--R",         "Temporal--Sup--L",   "Temporal--Pole--Sup--R",
"Temporal--Pole--Mid--L","Cerebellum--9--L",    "Vermis--1--2",         
"Vermis--4--5",       "Vermis--10".    \\
\small 
\\
\noindent The list of the regions of interest associated with the active set of MFG-Lasso when the response value is ADHD Inattentive:\\ 
\scriptsize
"Frontal--Mid--Orb--L","Frontal--Mid--Orb--R","Frontal--Inf--Oper--L", 
"Frontal--Inf--Oper--R","Frontal--Inf--Tri--L","Frontal--Inf--Orb--L",  
"Frontal--Inf--Orb--R","Supp--Motor--Area--L","Frontal--Sup--Medial--L"
"Frontal--Sup--Medial--R","Frontal--Med--Orb--L","Rectus--L",           
"Cingulum--Ant--L",   "Cingulum--Post--L",  "Cingulum--Post--R",    
"ParaHippocampal--R","Amygdala--L",       "Calcarine--L",        
"Cuneus--L",         "Cuneus--R",         "Lingual--L",          
"Occipital--Inf--L",  "Occipital--Inf--R",  "Parietal--Sup--L",     
"Parietal--Inf--L",   "SupraMarginal--L",  "SupraMarginal--R",    
"Angular--L",        "Angular--R",        "Precuneus--L",        
"Precuneus--R",      "Paracentral--Lobule--L","Paracentral--Lobule--R"
"Heschl--R",         "Temporal--Sup--L",   "Temporal--Pole--Sup--R",
"Temporal--Pole--Mid--L","Cerebellum--4--5--R",  "Vermis--1--2",         
"Vermis--4--5",       "Vermis--10".      \\
\small 
\\
\noindent The list of the regions of interest associated with the active set of MFG-Lasso when the response value is ADHD Hyper/Impulsive:\\ 
\scriptsize
"Frontal--Mid--Orb--L","Frontal--Mid--Orb--R","Frontal--Inf--Oper--L", 
"Frontal--Inf--Oper--R","Frontal--Inf--Tri--L","Frontal--Inf--Orb--L",  
"Frontal--Inf--Orb--R","Rolandic--Oper--R",  "Supp--Motor--Area--L",  
"Olfactory--L",      "Frontal--Sup--Medial--L","Frontal--Sup--Medial--R"
"Frontal--Med--Orb--L","Frontal--Med--Orb--R","Rectus--L",           
"Rectus--R",         "Cingulum--Ant--L",   "Cingulum--Mid--L",     
"Cingulum--Post--L",  "ParaHippocampal--R","Amygdala--L",         
"Amygdala--R",       "Calcarine--L",      "Cuneus--L",           
"Cuneus--R",         "Occipital--Sup--R",  "Occipital--Mid--R",    
"Occipital--Inf--L",  "Occipital--Inf--R",  "Parietal--Sup--L",     
"Parietal--Inf--L",   "Parietal--Inf--R",   "SupraMarginal--L",    
"SupraMarginal--R",  "Angular--L",        "Angular--R",          
"Putamen--L",        "Pallidum--R",       "Heschl--L",           
"Heschl--R",         "Temporal--Sup--L",   "Temporal--Pole--Sup--R",
"Temporal--Pole--Mid--L","Temporal--Pole--Mid--R","Cerebellum--3--R",      
"Cerebellum--4--5--R",  "Cerebellum--9--L",    "Vermis--1--2",         
"Vermis--3",         "Vermis--4--5",       "Vermis--6",           
"Vermis--7",         "Vermis--10".  
\\

\noindent
\small{The list of the regions that are associated with IQ but not with ADHD by the MFG-Lasso:}\\
\scriptsize "Frontal--Inf--Tri--R","Rolandic--Oper--R","Olfactory--R",   "Frontal--Med--Orb--R","Cingulum--Post--R", 
"Amygdala--R",    "Calcarine--R",   "Lingual--L",     "Occipital--Sup--L","Occipital--Sup--R", 
"Occipital--Mid--R","Parietal--Inf--R","Pallidum--R",    "Cerebellum--3--L", "Cerebellum--3--R",   
"Vermis--3",      "Vermis--6",      "Vermis--9", 
 \\

\small
\noindent
The list of the regions that are associated with ADHD but not with  IQ by the MFG-Lasso:\\
\scriptsize
"Frontal--Mid--L",    "Frontal--Sup--Medial--R","ParaHippocampal--R","Parietal--Inf--L",     
"Temporal--Pole--Sup--R","Cerebellum--9--L".
\normalfont

\vskip 0.2in

\end{document}